\documentclass[twocolumn,groupedaddress,preprintnumbers,amsmath,amssymb,aps,pre,floatfix]{revtex4}

\usepackage{xcolor}
\usepackage{layouts} 
\usepackage{dcolumn}
\usepackage[]{graphicx}
\usepackage{float} 

\usepackage[caption=false]{subfig} 

\usepackage{soul} 

\begin{document}
\title{On the choice of diameters in a polydisperse model glassformer:\\ deterministic or stochastic?}
\author{Niklas K\"uchler}
\affiliation{Institut f\"ur Theoretische Physik II: Weiche Materie, 
Heinrich-Heine-Universit\"at D\"usseldorf, Universit\"atsstra\ss e 1, 
40225 D\"usseldorf, Germany}

\author{J\"urgen Horbach}
\affiliation{Institut f\"ur Theoretische Physik II: Weiche Materie,
Heinrich-Heine-Universit\"at D\"usseldorf, Universit\"atsstra\ss e 1, 
40225 D\"usseldorf, Germany}

\date{\today}

\begin{abstract}
In particle-based computer simulations of polydisperse glassforming
systems, the particle diameters $\sigma = \sigma_1, \dots, \sigma_N$
of a system with $N$ particles are chosen with the intention to
approximate a desired distribution density $f$ with the corresponding
histogram.  One method to accomplish this is to draw each diameter
randomly from the density $f$. We refer to this stochastic scheme
as model $\mathcal{S}$.  Alternatively, one can apply a deterministic
method, assigning an appropriate set of $N$ values to the diameters.
We refer to this method as model $\mathcal{D}$. We show that
especially for the glassy dynamics at low temperatures it matters
whether one chooses model $\mathcal{S}$ or model $\mathcal{D}$.
Using molecular dynamics computer simulation, we investigate a
three-dimensional polydisperse non-additive soft-sphere system with
$f(s) \sim s^{-3}$. The Swap Monte Carlo method is employed to
obtain equilibrated samples at very low temperatures.  We show that
for model $\mathcal{S}$ the sample-to-sample fluctuations due to
the quenched disorder imposed by the diameters $\sigma$ can be
explained by an effective packing fraction.  Dynamic susceptibilities
in model $\mathcal{S}$ can be split into two terms: One that is of
thermal nature and can be identified with the susceptibility of
model $\mathcal{D}$, and another one originating from the disorder
in $\sigma$.  At low temperatures the latter contribution is the
dominating term in the dynamic susceptibility.
\end{abstract}

\maketitle

\section{Introduction}
Many of the colloidal systems that have been used to study the glass
transition are polydisperse \cite{gasser2009}. While monodisperse
colloidal fluids crystallize very easily, with the introduction of
a size polydispersity they become good glassformers \cite{vanmegen1993,
vanmegen1994, schope2006, schope2007, pham2001, pusey2009,
zaccarelli2015, brambilla2009}.  As a matter of fact, the degree
of polydispersity $\delta$, defined as the standard deviation of
the particle diameter divided by the mean particle diameter, may
strongly affect glassy dynamics. For example, for three-dimensional
hard-sphere colloids, it has been shown that for moderate polydispersity
$\delta < 10\%$ a dynamic freezing is typically seen for a packing
fraction $\phi_{\rm g} \approx 0.58$, while for $\delta \gtrsim
10\%$, the dynamics are more heterogeneous with the large particles
undergoing a glass transition at $\phi_{\rm g}$ while the small
particles are still mobile (note that this result is dependent on
the distribution of particle diameters) \cite{zaccarelli2015}.  An
interesting finding regarding the effect of polydispersity on the
dynamics has been reported in a simulation study of a two-dimensional
Lennard-Jones model \cite{klochko2020}. Here, Klochko {\it et
al.}~show that polydispersity is associated with composition
fluctuations that, even well above the glass-transition temperature,
lead to a two-step relaxation of the dynamic structure factor at
low wavenumbers and a long-time tail in the time-dependent heat
capacity. These examples demonstrate that polydispersity and the
specific distribution of particle diameters may strongly affect the
static and dynamic properties of glassforming fluids.

In a particle-based computer simulation, one can assign to each
particle $i$ a ``diameter'' $\sigma_i$. Note that in the following
the diameter of a particle does not refer to the geometric diameter
of a hard sphere, but in a more general sense it is a parameter
with the dimension of a length that appears in the interaction
potential between soft spheres (see below). To realize a polydisperse
system in the simulation of an $N$ particle system, one selects the
$N$ particle diameters to approximate a desired distribution density
$f(\sigma)$ with the corresponding histogram. Here, two approaches
have been used in previous simulation studies. In a stochastic
method, referred to as model $\mathcal{S}$ in the following, one
uses random numbers to independently draw each diameter $\sigma_i$
from the distribution $f$.  As a consequence, one obtains a
``configuration'' of particle diameters that differs from sample
to sample. Alternatively, to avoid this disorder, one can choose
the $N$ diameters in a deterministic manner, i.e.~one defines a map
$(f,\,N) \mapsto (\sigma_1,\dots,\sigma_N)$, which uniquely determines
$N$ diameter values. In the following, we refer to this approach
as model $\mathcal{D}$. The diameters in model $\mathcal{D}$ should
be selected such that in the limit $N\to\infty$ the histogram of
diameters converges to $f$ as being the case for model $\mathcal{S}$.
Unlike model $\mathcal{S}$, each sample of size $N$ of model
$\mathcal{D}$ has exactly the same realization of particle diameters.

Recent simulation studies on polydisperse glassformers have either
used model $\mathcal{S}$ (see, e.g., Refs.~\cite{zaccarelli2015,
klochko2020, leocmach2013, ingebrigtsen2015, ingebrigtsen2021,
ninarello2017, guiselin2020overlap, vaibhav2022, lamp2022}) or model
$\mathcal{D}$ schemes (see, e.g., Refs.~\cite{voigtmann2009,
weysser2010, santen2001liquid}).  However, a systematic study is
lacking where both approaches are compared.  This is especially
important when one considers states of glassforming liquids at very
low temperatures (or high packing fractions) where dynamical
heterogeneities are a dominant feature of structural relaxation.
For polydisperse systems, such deeply supercooled liquid states
have only recently become accessible in computer simulations, using
the Swap Monte Carlo technique \cite{tsai1978structure, grigera2001fast}.
For these states, the additional sample-to-sample fluctuations in
model $\mathcal{S}$ are expected to strongly affect static and
dynamic fluctuations in the system, as quantified by appropriate
susceptibilities.

In this work, we compare a model $\mathcal{S}$ to a model $\mathcal{D}$
approach for a polydisperse glassformer, using molecular dynamics
(MD) computer simulation in combination with the Swap Monte Carlo
(SWAP) technique. This hybrid scheme allows to equilibrate samples
at very low temperatures far below the critical temperature of mode
coupling theory. We analyze static and dynamic susceptibilities and
their dependence on temperature $T$ and system size $N$, keeping
the number density constant. We show that in the thermodynamic
limit, $N\to \infty$, the sample-to-sample fluctuations of model
$\mathcal{S}$ lead to a \textit{finite} static disorder susceptibility
of extensive observables.  This result is numerically shown for the
potential energy. Moreover, we analyze fluctuations of a time-dependent
overlap correlation function $Q(t)$ via a dynamic susceptibility
$\chi(t)$. At low temperatures, $\chi$ in model $\mathcal{S}$ is
strongly enhanced when compared to the one in model $\mathcal{D}$.
This finding indicates that it is crucial to carefully analyze the
disorder due to size polydispersity when one uses a model $\mathcal{S}$
approach.

In the next section \ref{sec:models}, we introduce the model for a
polydisperse soft-sphere system and define the models $\mathcal{S}$
and $\mathcal{D}$. The main details of the simulations are given
in Sec.~\ref{sec:simulation_details}. Then, Sec.~\ref{sec:thermodynamics}
is devoted to the analysis of static fluctuations of the potential
energy.  Here, we discuss in detail thermal fluctuations in terms
of the specific heat $C_V(T)$ and static sample-to-sample fluctuations
by a disorder susceptibility. In Sec.~\ref{sec:structural_dynamics},
dynamic fluctuations of the overlap function $Q(t)$ are investigated.
Finally, in Sec.~\ref{sec:conclusions}, we summarize and draw
conclusions.

\section{Polydisperse model system and choice of diameters}
\label{sec:models} 
{\it Particle interactions.} As a model glassformer, we consider a
polydisperse non-additive soft-sphere system of $N$ particles in
three dimensions. This model has been proposed by Ninarello {\it
et al.}~\cite{ninarello2017}.  The particles are placed in a cubic
box of volume $V=L^3$, where $L$ is the linear dimension of the
box.  Periodic boundary conditions are imposed in the three spatial
directions.  The particles have identical masses $m$ and their
positions and velocities are denoted by ${\bf r}_i$ and ${\bf v}_i$,
$i=1, \dots, N$, respectively. The time evolution of the system is
given by Hamilton's equations of motion with Hamiltonian $H = K +
U$. Here, $K = \sum_{i=1}^N {\bf p}_i^2/m$ is the total kinetic
energy and ${\bf p}_i = m {\bf v}_i$ the momentum of particle $i$.
Interactions between the particles are pairwise such that the total
potential energy $U$ can be written as
\begin{equation} 
U = \sum_{i=1}^{N-1} \sum_{j>i}^N
u(r_{ij}/\sigma_{ij}) \, . 
\label{eq_potentialenergy}
\end{equation} 
Here the argument of the interaction potential $u$ is
$x=r_{ij}/\sigma_{ij}$, where $r_{ij} = |\vec{r}_i - \vec{r}_j|$
denotes the absolute value of the distance vector between particles
$i$ and $j$. The parameter $\sigma_{ij}$ is related to the ``diameters''
$\sigma_i$ and $\sigma_j$, respectively, as specified below. The
pair potential $u$ is given by
\begin{equation} 
u(x) = u_0 \left(x^{-12} + c_0 + c_2
x^2 + c_4 x^4 \right)
\, \Theta(x_c - x) \, ,
\label{eq:U_pair} 
\end{equation} 
where the Heaviside step function $\Theta$ introduces a dimensionless
cutoff $x_c = 1.25$. The unit of energy is defined by $u_0$.  The
constants $c_0=-28 /x_c^{12}$, $c_2=48/x_c^{14}$, and $c_4=-21/
x_c^{16}$ ensure continuity of $u$ at $x_c$ up to the second
derivative.

We consider a polydisperse system, i.e.~each particle is allowed
to have a different diameter $\sigma_i$. In the following, lengths
are given in units of the mean diameter $\bar{\sigma}$, to be
specified below. A non-additivity of the particle diameters is
imposed in the sense that
\begin{equation}
\sigma_{ij} = \frac{\sigma_i + \sigma_j}{2}
\left( 1 - 0.2 |\sigma_i - \sigma_j| \right) \, .
\label{eq:non_additivity}
\end{equation}
This non-additivity has been introduced to suppress crystallization
\cite{ninarello2017} which is in fact provided down to temperatures
far below the critical temperature of mode coupling theory.

{\it Choice of particle diameters.} The diameters $\sigma_i$ of the
particles are chosen according to two different protocols.  In model
$\mathcal{S}$, each diameter is drawn independently from the same
probability density $f(\sigma)$. In model $\mathcal{D}$, the diameters
for a system of size $N$ are chosen in a deterministic manner such
that their histogram approximates $f$ in the limit $N\to \infty$.
As in Ref.~\cite{ninarello2017}, we consider a function $f(\sigma)\sim
\sigma^{-3}$.  In the case of an additive hard-sphere system, this
probability density ensures that within each diameter interval of
constant width the same volume is occupied by the spheres.

\textit{Model $\mathcal{S}$.} For model $\mathcal{S}$, particle
diameters $\sigma_i$ are independently and identically distributed,
each according to the same distribution density
\begin{equation}
f(\sigma) = A \sigma^{-3} 
\mathbf{1}_{[\sigma_{\rm m}, \sigma_{\rm M}]}(\sigma) \,.
\label{eq_fsigma}
\end{equation}
Here $\mathbf{1}_B(\sigma)$ denotes the indicator function, being
one if $\sigma \in B$ and $0$ otherwise. The normalization $\int
f(\sigma)\;\text{d}\sigma = 1$ is provided by the choice $A = 2 /
(\sigma_{\rm m}^{-2} - \sigma_{\rm M}^{-2})$. We define the unit
of length as the expectation value of the diameter,
\begin{equation}
\bar{\sigma} = \int \sigma f(\sigma)  ~\text{d}\sigma \, ,
\label{eq:sigma_expectation}
\end{equation}
which implies $\sigma_{\rm M} = \sigma_{\rm m} /(2 \sigma_{\rm m}
- 1)$. We set the lower diameter bound to $\sigma_{\rm m} = 29/40
= 0.725$. Thus, the upper bound is given by $\sigma_{\rm M} = 29/18
= 1.6\overline{1}$ and the amplitude in Eq.~(\ref{eq_fsigma}) is
$A = 29/22 = 1.3\overline{18}$.  Note that the ratio $\sigma_{\rm
m}/\sigma_{\rm M} = 20/9 = 2.\overline{2}$, chosen in this work,
deviates by less than 0.24\% from the values $2.219$ and $2.217$
reported in Refs.~\cite{ninarello2017} and
\cite{RFIM_in_glassforming_liquid}, respectively. The degree of
polydispersity $\delta$ can be defined via the equation $\delta^2
= \int (s - \bar{\sigma})^2 f(s) \text{d}s/\bar{\sigma}^2$ and has
the value $\delta \approx 22.93\%$ in our case.

In practice, random numbers $\sigma$ following a distribution $f$
can be generated from a uniform distribution on the interval $[0,1]$
via the method of inversion of the cumulative distribution function
(CDF). The CDF is defined as
\begin{equation}
F(\sigma) = 
\int_{-\infty}^\sigma \, f(s) \;\text{d}s \, .
\label{eq:CDF_sigma}
\end{equation}
Its codomain is the interval $[0,1]$. Now the idea is to use a
uniform random number $Y \in [0,1]$ to select a point on the codomain
of $F$.  Then, via the inverse of the CDF, $F^{-1}: [0,1] \to
[\sigma_{\rm m}, \sigma_{\rm M}]$, one can map $Y$ to the number
\begin{equation}
\sigma = F^{-1}(Y) = \left(\frac{1}{\sigma^2} 
- \frac{2}{A} Y \right)^{-1/2} \, ,
\label{eq_sigmai}
\end{equation}
which follows the distribution $f$ as desired.

The empirical CDF, $F_N$, associated with a sample of $N$ diameter
values, reads
\begin{equation}
F_N(\sigma) = N^{-1} \sum_{i=1}^N 
\mathbf{1}_{\left( -\infty, \sigma \right]}(\sigma_i). 
\label{eq:CDF_empirical_sigma}
\end{equation}
Since for model $\mathcal{S}$ the diameters $\sigma_i$ are independently
and identically distributed according to the CDF $F$, the following
relation holds for all $\sigma \in \mathbb{R}$,
\begin{equation}
\lim_{N \to \infty} F_N^{\mathcal{S}}(\sigma) 
\stackrel{\textit{almost surely}}{=} F(\sigma) \,.
\label{eq:CDF_S_convergence}
\end{equation}
This follows from the strong law of large numbers.

\textit{Additive packing fraction.} To a hard-sphere sample with
particle diameters $\sigma_i$, $i=1, \dots, N$, one can assign the
additive hard-sphere packing fraction
\begin{equation}
\phi_\mathrm{hs} = \frac{1}{V} \sum_{i=1}^{N} \frac{\pi}{6} \sigma_i^3.
\label{eq_phihs}
\end{equation}
For model $\mathcal{S}$, the value of $\phi_{\rm hs}$ fluctuates
among independent samples of size $N$ around the expectation value
\begin{equation}
\phi^\infty_\mathrm{hs} := \mathrm{E}^\mathcal{S}[\phi_{\rm hs}] = 
\frac{\pi n}{6} A \left( \sigma_{\rm M} - \sigma_{\rm m} \right) 
\approx 0.612 \,.
\label{eq:expectation_packing_fraction}
\end{equation}
Here $n=N/V$ is the number density and the expectation
$\mathrm{E}^\mathcal{S}[\,.\,]$ is calculated with respect to the
diameter distribution $\prod_{i=1}^{N} f(\sigma_i)$ on the global
diameter space.  The variance of $\phi_{\rm hs}$ can be written as
\begin{equation}
\mathrm{Var}^\mathcal{S}(\phi_{\rm hs}) =
N^{-1} \left(\frac{\pi n}{6}\right)^2 
\mathrm{Var}^\mathcal{S}(\sigma^3) \, , 
\label{eq:fluctuations_phi}
\end{equation}
where $\mathrm{Var}^\mathcal{S}(\sigma^3)$ is the variance of
$\sigma_i^3$ for a single particle. The fluctuations
$\mathrm{Var}^\mathcal{S}(\phi_{\rm hs}) \propto N^{-1}$ vanish for
$N\to\infty$.  Beyond that, the disorder susceptibility
\begin{equation}
\chi_\mathrm{dis}^\mathcal{S}[\phi_\mathrm{hs}] = 
N \mathrm{Var}^\mathcal{S}(\phi_{\rm hs}) 
= \textit{Const} > 0
\end{equation}
is constant and finite for model $\mathcal{S}$.  In
Sec.~\ref{Section_disorder_fluctuations}, the disorder fluctuations
for model $\mathcal{S}$ will be discussed and analyzed in more
depth.

Note that $\phi_\mathrm{hs}$ is not an appropriate measure for a
non-additive polydisperse model that we use in our work.  Therefore,
later on, we will define an effective packing fraction $\phi_\mathrm{eff}$
to account for non-additive particle interactions.
\begin{figure}
\includegraphics{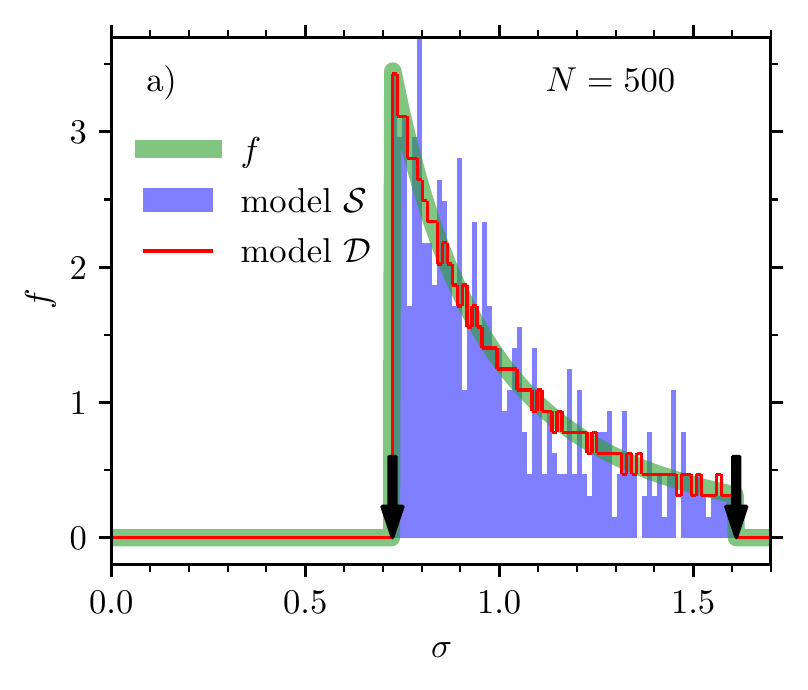}
\includegraphics{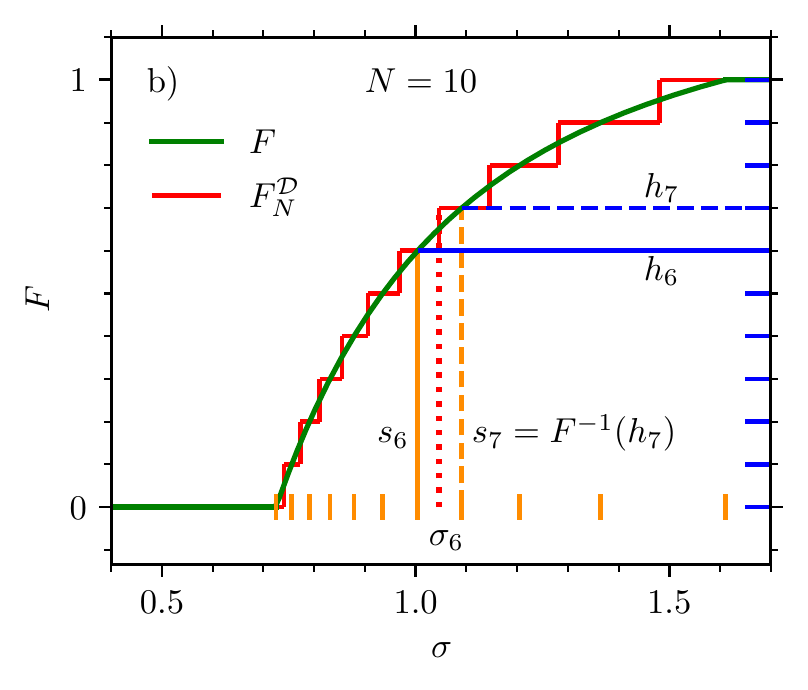}
\caption{a) Histogram of $N = 500$ particle diameters $\sigma_i$
of models $\mathcal{S}$ (blue) and $\mathcal{D}$ (red), respectively.
For model $\mathcal{S}$ a single realization is shown, where each
$\sigma_i$ is drawn independently from the density $f(\sigma)$
(green). In both histograms $70$ bins are used. The vertical arrows
indicate the minimum and maximum diameters, $\sigma_{\rm m}$ and
$\sigma_{\rm M}$, respectively. b) Cumulative distribution function
(CDF) $F$ (green) and empirical CDF $F_N^\mathcal{D}$ for model
$\mathcal{D}$ (red) as a function of diameter $\sigma$ for the
example $N=10$. The diameters $\sigma_i$ are constructed from
Eqs.~(\ref{eq_hi}-\ref{eq:poly_dispersity_diameters_D}), as graphically
illustrated for $\sigma_6$. \label{fig1}}
\end{figure}

\textit{Model $\mathcal{D}$.} For model $\mathcal{D}$, we also use
the CDF $F$ to obtain the particle diameters $\sigma_i$, $i= 1,
\dots, N$, but now we generate them in a deterministic manner. Our
upcoming construction will satisfy the following three conditions:
\begin{enumerate}
\item The construction is deterministic. The system size $N$ uniquely 
defines the diameters,
\begin{equation}
N~\mapsto~\sigma_1, \dots, \sigma_N.
\end{equation}
\item Convergence: The empirical CDF $F_N^\mathcal{D}$ approximates 
$F$. The convergence is uniform,
\begin{equation}
\lim_{N \to \infty} F_N^{\mathcal{D}} 
\stackrel{\textit{uniform}}{=} F \,.
\end{equation}
Thus the models $\mathcal{S}$ and $\mathcal{D}$ are consistent.
\item Constraint: For a given one-particle property $\theta(\sigma)$ 
of the diameter, the following constraint is fulfilled:
\begin{equation}
\frac{1}{N} \sum_{i=1}^{N} \theta(\sigma_i) = 
\mathrm{E}^\mathcal{S}[\,\theta\,].
\label{eq:constraint_model_D}
\end{equation}
This means that the empirical mean of the function $\theta(\sigma_i)$
equals the corresponding expectation $E^\mathcal{S}[\,\theta(\sigma_i)\,]$
in model $\mathcal{S}$. To ensure this, $\theta$ is required to be
a strictly monotonic function in $\sigma$.
\end{enumerate}
For our work, we use $\theta(\sigma) = \frac{\pi}{6} \sigma^3$,
inspired by the additive hard-sphere packing fraction,
cf.~Eq.~(\ref{eq_phihs}).  Here, Eq.~(\ref{eq:constraint_model_D})
ensures that $\phi_\mathrm{hs}$ has the same value for any $N$,
\begin{equation}
\phi_\mathrm{hs}^\mathcal{D}
= \mathrm{E}^\mathcal{S}[\phi_\mathrm{hs}]  
\equiv \phi_\mathrm{hs}^\infty.
\end{equation}

So, how do we define the $N$ diameters $\sigma_i$ in the framework
of model $\mathcal{D}$? First, we introduce $N+1$ equidistant nodes
along the the codomain of $F$,
\begin{align}
h_i = i/N, && i=0, \dots, N.
\label{eq_hi}
\end{align}
Their pre-images $s_i$ are found on the domain of $F$,
\begin{equation}
s_i = F^{-1}(h_i)
\label{eq:poly_dispersity_diameters_D_si} \, .
\end{equation}
We then define particle diameters $\sigma_i$, $i=1, \dots, N$, via
\begin{equation}
\theta(\sigma_i) = N \int_{s_{i-1}}^{s_i} 
\theta(\sigma) f(\sigma)\,\mathrm{d}\sigma \, .
\label{eq:poly_dispersity_diameters_D}  
\end{equation}
Since $\theta$ is assumed to be strictly monotonic, its inverse
$\theta^{-1}$ exists and $\sigma_i$ is uniquely defined by
Eq.~(\ref{eq:poly_dispersity_diameters_D}).  By summing over $i$
the constraint Eq.~(\ref{eq:constraint_model_D}) is fulfilled.  The
proof of the uniform convergence $\lim_{N \to \infty} F_N^{\mathcal{D}}
= F$ is presented in Appendix~\ref{app:Convergence_CDF_D}.  Note
the analytical nature of the convergence for model $\mathcal{D}$
in contrast to the stochastic one for model $\mathcal{S}$,
cf.~Eq.~(\ref{eq:CDF_S_convergence}).

Equation~(\ref{eq:poly_dispersity_diameters_D}) with the choice
$\theta(\sigma) = \frac{\pi}{6} \sigma^3$ is a sensible constraint
for an additive hard-sphere system. For our non-additive soft-sphere
system it is a minor tweak and not an essential condition. Another
reasonable choice would be $\theta(\sigma) = \sigma$, which ensures
that the empirical mean of the diameters exactly equals the unit
of length $\bar{\sigma}$. Alternatively, one could ignore the
constraint Eq.~(\ref{eq:constraint_model_D}) and thus also
Eq.~(\ref{eq:poly_dispersity_diameters_D}) entirely and define
$\sigma_i = s_i$ via Eq.~(\ref{eq:poly_dispersity_diameters_D_si})
-- note that one obtains $N+1$ diameters in this case. The latter
approach was used in Ref.~\cite{santen2001liquid}. We expect that
all these options are equivalent in the limit $N \to \infty$.

Figure \ref{fig1}a illustrates the distribution of diameters for
the models $\mathcal{S}$ and $\mathcal{D}$. In each case, we show
one histogram for $N=500$ particles, in comparison to the distribution
density $f$. For a meaningful comparison, we have chosen the same
number of 70 bins for both histograms. Since model $\mathcal{S}$
is of stochastic nature, we show the histogram for a single realization
of diameters. In contrast, for model $\mathcal{D}$ the histogram
at a given $N$ and bin number is uniquely defined (assuming an
equidistant placement of bins on $[\sigma_\mathrm{m},\sigma_\mathrm{M}]$).
The fluctuations around $f$ for model $\mathcal{S}$ appear to be
larger than  for $\mathcal{D}$.  In the following paragraph ``Order
of convergence'', we put this finding on an analytical basis.

Figure~\ref{fig1}b illustrates the construction of diameters
$\sigma_i$ for model $\mathcal{D}$, based on the CDF $F$, for a
small sample size $N=10$.  For the resulting diameters the empirical
CDF $F_N^\mathcal{D}$ is shown.

{\it Order of convergence.} Having established the convergence
$\lim_{N\to\infty} F_N = F$ for models $\mathcal{S}$ and $\mathcal{D}$,
we now compare their order of convergence. To this end, we calculate
$\Delta F$, defined as the square-root of the mean squared deviation
between $F_N$ and $F$,
\begin{equation}
\Delta F = (\mathrm{E}[(F_N - F)^2])^{1/2} \, .
\label{eq:Delta_F}
\end{equation}
Here, $\mathrm{E}[\,.\,]$ refers to the expectation with respect
to the global diameter distribution. For model $\mathcal{D}$, the
expectation $\mathrm{E}[\,.\,]$ is trivial and we obtain $\Delta
F^\mathcal{D} = |F_N^\mathcal{D} - F|$. As shown in the Appendices
\ref{app:Convergence_CDF_D} and \ref{app:CDF_order_of_convergence},
the results for model $\mathcal{D}$ and $\mathcal{S}$ are respectively
\begin{align}
\Delta F^\mathcal{D} &\leq N^{-1}\,,\\
\Delta F^\mathcal{S} &= \left((F(1-F)\right)^{1/2} N^{-1/2}\,.
\end{align}
This means that the order of convergence for model $\mathcal{D}$
is at least $1$, in contrast to model $\mathcal{S}$ where the order
is only $1/2$.  In this aspect, model $\mathcal{D}$ is superior to
model $\mathcal{S}$, since its diameter distribution approaches the
thermodynamic limit faster.  Numerically, from the equations above,
one has $\max_\sigma \Delta F^\mathcal{D} \leq \max_\sigma \Delta
F^\mathcal{S}$ already for $N \geq 4$.

\section{Simulation details}
\label{sec:simulation_details}
Depending on the protocols introduced below, different particle-based
simulation techniques are used, among which are molecular dynamics
(MD) simulations, the Swap Monte Carlo (SWAP) method, and the
coupling of the system to a Lowe-Andersen thermostat (LA).

\begin{figure}
\includegraphics{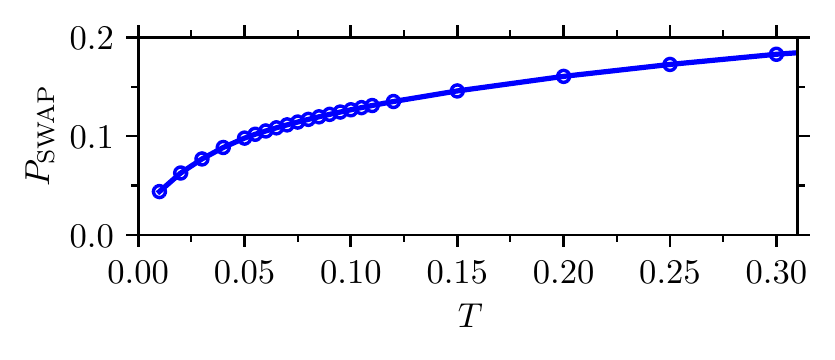}
\caption{Acceptance rate $P_\mathrm{SWAP}$ of diameter exchange
trials as a function of temperature $T$.  \label{fig2}}
\end{figure}

In the MD simulations, Newton's equations of motion are numerically
integrated via the velocity form of the Verlet algorithm, using a
time step of $\Delta t = 0.01\,t_0$ (with $t_0 = \bar{\sigma}
\sqrt{m/u_0}$ setting the unit of time in the following). We employ
the SWAP method in combination with MD
simulation~\cite{berthier2019efficient}. To this end, every 25 MD
steps, $N$ trial SWAP moves are performed. In a single SWAP move,
a particle pair $(i,j)$ is randomly selected, followed by the attempt
to exchange their diameters $(\sigma_i,\sigma_j)$ according to a
Metropolis criterion. The probability $P_\mathrm{SWAP}$ to accept
a SWAP trial as a function of $T$ is shown in Fig.~\ref{fig2}. It
indicates that even deep in the glassy state (far below the glass-transition 
temperature $T_{\rm g}^\mathrm{SWAP} \approx 0.06$, which
we will define later on), the acceptance rate for a SWAP move is
still $\gtrsim 4\%$ for $T \geq 0.01$. The latter is the lowest
temperature shown here.

During the equilibration protocols, in each step, we couple the
system to a Lowe-Andersen
thermostat~\cite{Lowe_Andersen_T_different_masses} for identical
masses $m$ to reach a target temperature $T$: For each particle
pair $(i,j)$ closer than a cutoff $R_\mathrm{T}$ and with a probability
$\Gamma \Delta t$ new velocities are generated as
\begin{equation}
{\bf v}_{i/j}^\mathrm{new} = {\bf v}_{i/j} 
\pm \frac{1}{2}\left( \zeta \sqrt{\frac{2 k_B T}{m}} - 
({\bf v}_i -{\bf v}_j)\cdot \hat{\bf r}_{ij} \right) \hat{\bf r}_{ij},
\end{equation}
where $\hat{\bf r}_{ij}  = \mathbf{r}_{ij}/|\mathbf{r}_{ij}|$ and
$\zeta$ is a normally distributed variable with expectation value
of $0$ and variance of $1$. This means that only the component of
the relative velocity parallel to $\hat{\bf r}_{ij}$ is thermalized,
preserving the momentum as well as the angular momentum. We choose
$R_\mathrm{T} = x_c$ and $\Gamma = 4$.

Both for model $\mathcal{S}$ and model $\mathcal{D}$, we consider
different system sizes $N = 256$, $500$, $1000$, $2048$, $4000$,
and $8000$ particles at different temperatures $T$, respectively.
In each case, we prepare $60$ independent configurations as follows:
The initial positions are given by a face-centered-cubic lattice
(with cavities in case that $N \neq 4 k^3$ for all integers $k$),
while the initial velocities have a random orientation with a
constant absolute value according to a high temperature $T = 5$.
The total momentum is set to $\bf{0}$ by subtracting $\sum_{i} {\bf
v}_i /N$ from the velocity of each particle.  The initial crystal
is melted for a simulation time $t_\text{max} = 2000$ with $\Delta
t = 0.001$, applying both the SWAP Monte Carlo and the LA thermostat.
Then we cool the sample to $T = 0.3$ for the same duration, followed
by a run with $\Delta t = 0.01$ over the time $t_\text{max} = 10^5$
to fully equilibrate the sample at the target temperature $T$. After
that we switch off SWAP (to ensure that the mean energy remains
constant in the following) and measure a time series $H(t_j)$ of
the total energy over a time span of $0.75\,t_\text{max}$. Then we
calculate the corresponding mean $H_\mathrm{av}$ and the standard
deviation $\mathrm{sd}(H)$, and as soon as the condition $|H(t) -
H_\mathrm{av}| < 0.01\,\mathrm{sd}(H)$ is met, we switch off the
LA thermostat and perform a microcanonical $NVE$ simulation for the
remaining time up to $t=t_\text{max}$. This procedure reduces
fluctuations in the final temperature $T$ for subsequent $NVE$
production runs.

For the analysis that we present in the following, we mostly compare
$NVE$ with SWAP production runs (in both cases without the LA
thermostat).  Also, we perform MD production runs with the coupling
to the LA thermostat but without applying the SWAP, and accordingly
refer to these runs as the LA protocol. For all of these production
runs, the initial configurations are the final samples obtained
from the equilibration protocol described above.

For the LA thermostat and the SWAP Monte Carlo, pseudorandom numbers
are generated by the \textit{Mersenne Twister}
algorithm~\cite{matsumoto1998mersenne}. For each sample, a different
seed is chosen to ensure independent sequences.  For an observable
we eventually determine its $95\%$ confidence interval from its
empirical CDF, which is calculated via
Bootstrapping~\cite{efron1992bootstrap} with $1000$ repetitions.

\section{Static fluctuations}
\label{sec:thermodynamics}
In the following two subsections ``Thermal fluctuations'' and
``Disorder fluctuations'', we consider two kinds of fluctuations.
Thermal fluctuations quantify \textit{intrinsic fluctuations of
phase-space variables for a given diameter configuration}. These
intrinsic observables are expected to coincide for both models
$\mathcal{S}$ and $\mathcal{D}$, provided that  $N$ is sufficiently
large. As an example, we study thermal energy fluctuations, as
quantified by the specific heat (here, numerical results are only
shown for model $\mathcal{D}$). Below, we use this quantity to
determine the glass-transition temperatures for the different
dynamics.

In model $\mathcal{S}$, the dependence of thermally averaged
observables on the diameter configuration leads to sample-to-sample
fluctuations that are absent in model $\mathcal{D}$. We measure
these fluctuations in terms of a disorder susceptibility, exemplified
via the potential energy.

\subsection{Thermal fluctuations}
Let us consider an $N$ particle sample of our system. An observable
$O$ that characterizes the state of this sample depends in general
on the particle coordinates $r=({\bf r}_1, \dots, {\bf r}_N$), the
momenta $p=({\bf p}_1, \dots, {\bf p}_N)$, and the particle diameters
$\sigma = (\sigma_1, \dots, \sigma_N)$. When we denote the phase-space 
configuration by $q=(r,p)$, we can write the observable as
$O = O(q,\sigma)$. Its thermal average can be expressed as
\begin{equation}
\langle O \rangle(\sigma) = \mathrm{E}(O|\sigma) = 
\int O(q,\sigma) \rho(q|\sigma)~\mathrm{d}q \, ,
\end{equation}
where $\rho( q | \sigma)$ is a \textit{conditional} phase-space
density.  In the case of the canonical $NVT$ ensemble, it is given
by
\begin{equation}
\rho( q | \sigma) = Z^{-1} \exp( - H(q|\sigma)/(k_B T) )
\label{eq_condpsd}
\end{equation}
with $Z = \int \exp( - H(q|\sigma)/(k_B T) ) ~\mathrm{d}q$ the
partition function and $H=K+U$ the Hamiltonian, cf.~Sec.~\ref{sec:models}.

In the simulations, we compute $\langle O \rangle(\sigma)$ via the
average of an equidistant time sequence $q(t_i)$ (with $\#t_i =
5000$) over a time window $t_{\rm max} = 10^5$. This approach is
valid for an ergodic system - by definition - in case sufficient
sampling is ensured. Then, the result \textit{does not} depend on
the initial condition $q(0)$. However, it \textit{does} depend on
the realization of $\sigma$ and, of course, the ensemble parameters,
e.g.~the temperature $T$.

Thermal fluctuations of the observable $O$ can be quantified in
terms of the thermal susceptibility 
\begin{equation} \chi_\mathrm{thm}[O]
= \mathrm{Var}(O|\sigma)/N = \langle O^2 - \langle O \rangle^2
\rangle  / N \, .  
\label{eq:thermal_sus} 
\end{equation} 
Here the variance $\mathrm{Var}(\,.\,)$ is calculated according to
the phase-space density~(\ref{eq_condpsd}). The normalization for
$\chi_\mathrm{thm}$ is chosen such that for an extensive observable
$O$ we expect finite values for $\lim_{N \to \infty} \chi_\mathrm{thm}[O]$.

An important quantity that is related to the thermal susceptibility
of the potential energy $U$ is the excess specific heat at constant
volume, defined by
\begin{equation}
C_V = 
\frac{1}{N} \frac{\partial \langle U \rangle}{\partial T} \, . 
\label{eq:CV_derivative}
\end{equation}
In the canonical $NVT$ ensemble, the relation between $C_V$ and the
thermal susceptibility $\chi_\mathrm{thm}^{NVT}[U]$ is
\begin{equation}
C_V =
\chi_\mathrm{thm}^{NVT}[U]/T^2  
\label{eq:CV_fluctuations_NVT} \, .
\end{equation}
This formula can be converted to the microcanonical $NVE$ ensemble
to obtain \cite{lebowitz1967ensemble}
\begin{equation}
C_V =
\frac{\chi_\mathrm{thm}^{NVE}[U] } 
{T^2 - (2/3)\chi_\mathrm{thm}^{NVE}[U]} \, . 
\label{eq:CV_fluctuations_NVE}
\end{equation}
\begin{figure}
\includegraphics{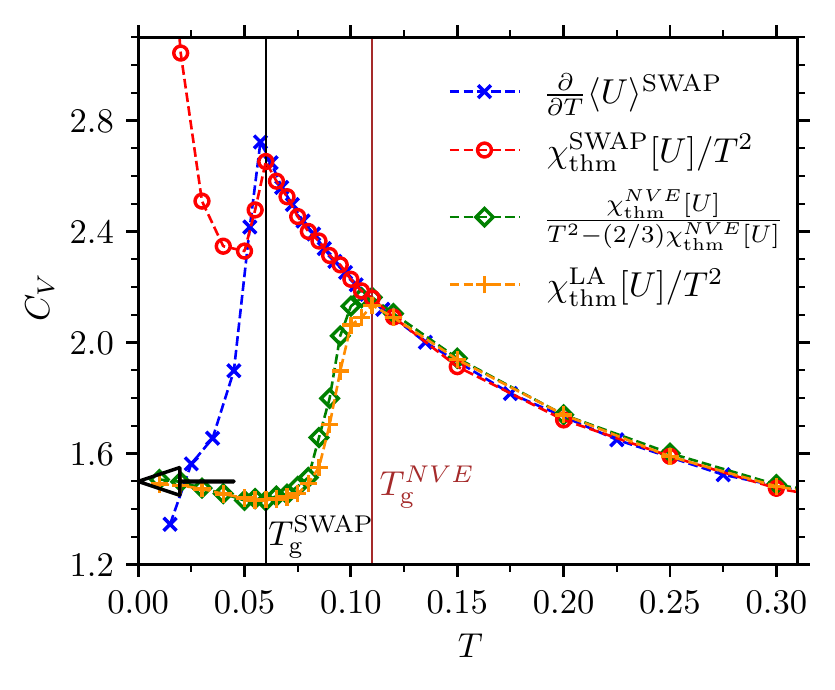}
\caption{Specific heat $C_V$ as a function of temperature $T$ for
model $\mathcal{D}$ with $N=2048$ particles. The solid lines indicate
the glass-transition temperatures, corresponding to the microcanonical
MD simulations (green, $T_\mathrm{g}^{NVE} = 0.11$) and the simulations
with SWAP dynamics (blue and red, $T_\mathrm{g}^\mathrm{SWAP} =
0.06$).  Coupling to the LA thermostat but without SWAP is represented
by the orange line.  The black arrow indicates the Dulong-Petit
limit, $C_V = 3/2$. \label{fig3}}
\end{figure}
Figure~\ref{fig3} shows $C_V$ as a function of temperature $T$ for
the different dynamics, namely the microcanonical MD via
Eq.~(\ref{eq:CV_fluctuations_NVE}), the MD with SWAP using
Eqs.~(\ref{eq:CV_derivative}) and (\ref{eq:CV_fluctuations_NVT}),
and the MD with LA thermostat employing again
Eq.~(\ref{eq:CV_fluctuations_NVT}).

At high temperatures, $T\gtrsim 0.11$, the specific heat $C_V$ from
the different calculations is in perfect agreement.  Upon decreasing
$T$, one observes relatively sharp drops in $C_V$ for the microcanonical
$NVE$ and the SWAP dynamics.  The drops occur at the temperatures
$T_\mathrm{g}^{NVE} = 0.11$ and $T_\mathrm{g}^\mathrm{SWAP} = 0.06$,
respectively, and indicate the glass transition of the different
dynamics. These estimates of the glass-transition temperatures
$T_\mathrm{g}$ are consistent with those obtained from dynamic
correlation functions presented in Sec.~\ref{sec:structural_dynamics}.

Another conclusion that we can draw from Fig.~\ref{fig3} is that
fluctuations in $U$, as quantified by the $C_V$ from the SWAP
dynamics simulations, correctly reproduce those in the canonical
$NVT$ ensemble.  This can be inferred from the coincidence of the
blue and the red data points at temperatures $T >
T_\mathrm{g}^\mathrm{SWAP}$. For the $NVE$ dynamics at $T <
T_\mathrm{g}^{NVE}$, albeit using fully equilibrated samples as
initial configurations for $T > T_\mathrm{g}^\mathrm{SWAP}$,
relaxation times become too large to correctly resolve the fluctuations,
as quantified by $\chi_\mathrm{thm}^{NVE}[U]$. We underestimate
them within our finite simulation time and effectively measure a
frequency-dependent specific heat \cite{scheidler2001}. Thus, from
the monotonicity of Eq.~(\ref{eq:CV_fluctuations_NVE}), $C_V$ is
underestimated as well.  Furthermore, from the coincidence of the
green with the orange data points, corresponding to the NVE and LA
dynamics, respectively, we can conclude that the LA thermostat
correctly reproduces the fluctuations in the canonical $NVT$ ensemble.

For the $NVE$ as well as LA dynamics, we see the Dulong-Petit law,
i.e.~for $T \to 0$ the specific heat approaches the value $C_V=3/2$.
An exception to this finding are the results calculated from the
SWAP dynamics.  This can be understood by the fact that the SWAP
dynamics are associated with fluctuating particle diameters even
at very low temperatures; thus the resulting dynamics cannot be
described in terms of the harmonic approximation for a frozen solid.

\subsection{Disorder fluctuations}
\label{Section_disorder_fluctuations}
In model $\mathcal{S}$, the Hamiltonian $H(q|\sigma)$ is parameterized
by random variables $\sigma$ and this imposes a quenched disorder
onto the system.  This leads to fluctuations that can be quantified
in terms of a disorder susceptibility that we shall define and
analyze in this section.

To this end, we first introduce the diameter distribution density
for both models,
\begin{align}
g(\sigma) &= \begin{cases}
\Pi_{i=1}^N f(\sigma_i), &\mathrm{model}~\mathcal{S},\\
\Pi_{i=1}^N \delta_\mathrm{D}(\sigma_i - \sigma_i^\mathcal{D}),
&\mathrm{model}~\mathcal{D}\, ,
\end{cases} 
\label{eq:density_sigma_global}
\end{align}
where $\delta_\mathrm{D}$ denotes the Dirac delta function.

Let us consider a variable $B = B(\sigma)$. This could be a function
such as the additive hard-sphere packing fraction $\phi_{\mathrm{hs}}$
or the thermal average of a phase-space function at a given diameter
configuration $\sigma$, e.g.~$\langle U \rangle$. The disorder
average of $B$, denoted by $\overline{B}$, is the expectation value
of $B$ with respect to the distribution density $g$,
\begin{equation}
\overline{B} = \mathrm{E}(B) = 
\int B(\sigma) g(\sigma)~\mathrm{d}\sigma \, .
\end{equation}
Note that in our analysis below, disorder averages are calculated
by an average over all samples, i.e.~over $60$ realizations of
$\sigma$.

Fluctuations of an extensive quantity $B \sim N$ and its corresponding
``density'' $b = B/N$ can be measured by disorder susceptibilities,
defined as
\begin{align}
\chi_\mathrm{dis}[B] &=
\mathrm{Var}(B)/N =
\overline{ B^2 - \overline B^2 } / N \,,\\
\chi_\mathrm{dis}[b] &=
N \mathrm{Var}(b).
\label{eq:disorder_sus}
\end{align}
These two different definitions have to be applied for a meaningful
scaling, i.e.~to ensure $\chi_\mathrm{dis}[B] = \chi_\mathrm{dis}[b]$.
For model $\mathcal{D}$, we have $\chi_\mathrm{dis}^\mathcal{D}[B]
= 0$ for any $B$.  In contrast, for model $\mathcal{S}$ the variable
$B(\sigma)$ fluctuates from sample to sample as quantified by
$\chi_\mathrm{dis}[B]$. Here, in general,
$\lim_{N\to\infty}\chi_\mathrm{dis}[B] \neq 0$,  as exemplified by
the fluctuations of the additive packing fraction: In
Sec.~\ref{sec:models}, we showed
$\mathrm{Var}^\mathcal{S}(\phi_\mathrm{hs}) \propto 1/N$, and thus
we have $\chi^\mathcal{S}_\mathrm{dis}[\phi_\mathrm{hs}] =
\textit{Const} > 0$.

\begin{figure}
\centering 
\includegraphics[width=\columnwidth]{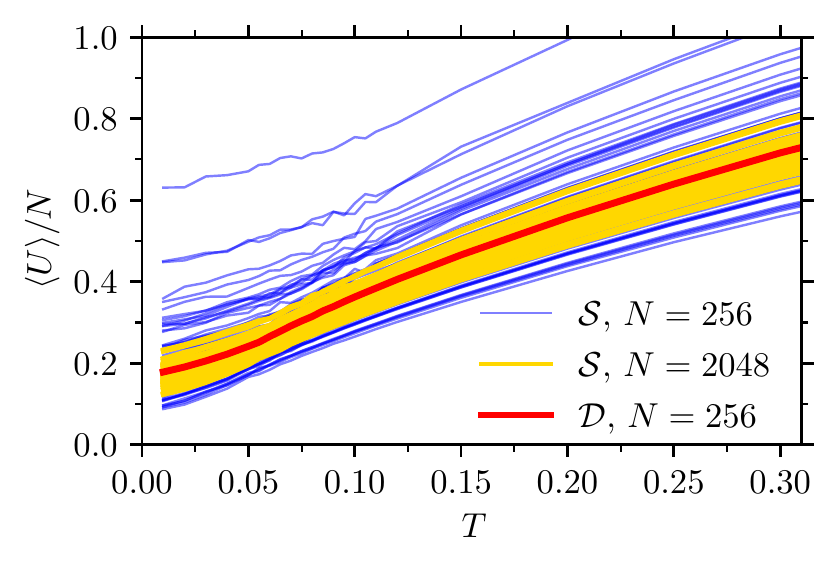}
\includegraphics[width=\columnwidth]{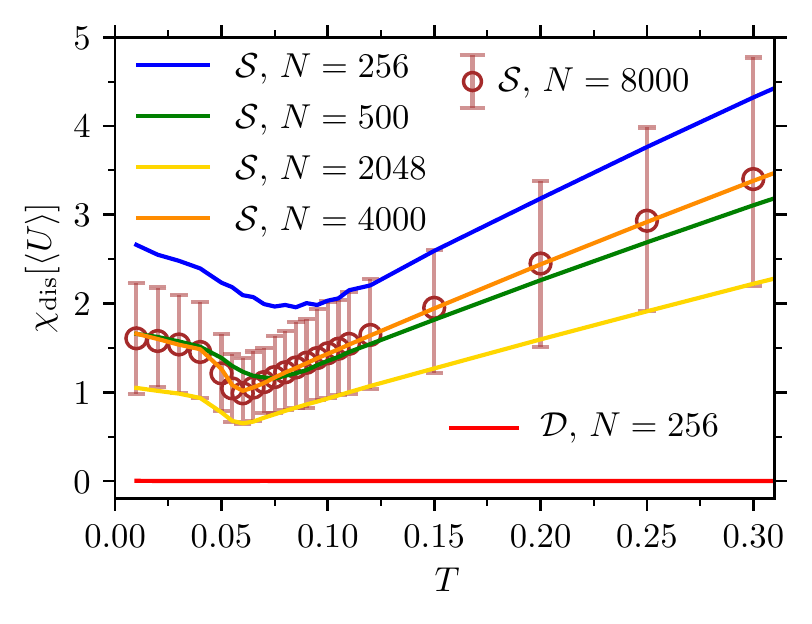}
\caption{a) Mean potential energy $\langle U \rangle(\sigma)$ as a
function of temperature $T$. For model $\mathcal{S}$, individual
curves for each of the 60 samples are shown for systems with $N=256$
(blue lines) and $N=2048$ (orange lines) and for model $\mathcal{D}$
for the system with $N=256$. b) Disorder susceptibility
$\chi_\mathrm{dis}[\langle U \rangle]$ for different values of $N$.
\label{fig4}}
\end{figure}
\textit{Potential energy.} Having introduced the disorder average
and susceptibility, we consider the variable $B(\sigma) = \langle
U \rangle(\sigma)$, corresponding to the thermal average of the
potential energy for a given sample with diameter configuration
$\sigma$.

In Figure~\ref{fig4}a the dependence of $\langle U \rangle(\sigma)$
on temperature $T$ is shown. For a given model and system size $N$,
we present $60$ curves corresponding to $60$ independent samples.
For model $\mathcal{S}$, results for $N=256$ and $2048$ are shown.
Here, the diameter configurations $\sigma$ vary among the samples
and thus, the potential energy fans out into various curves $\langle
U \rangle(T)$.  If we measure the fluctuations of the mean potential
energy per particle, $\langle U \rangle(\sigma)/N$, \textit{with
its variance}, the fluctuations decrease with increasing $N$, as
expected.  For model $\mathcal{D}$, we show the curves of $60$
independent samples at $N=256$; here, sample-to-sample fluctuations
are completely absent and all data collapse onto a single curve.

Figure \ref{fig4}b shows the disorder susceptibility
$\chi_\mathrm{dis}[\langle U \rangle]$ of model $\mathcal{S}$ for
different system sizes. As can be inferred from the figure, in a
non-monotonous manner, $\chi_\mathrm{dis}[\langle U \rangle]$ seems
to approach a finite temperature-dependent value in the limit $N\to
\infty$,
\begin{equation}
\lim_{N \to \infty} 
\chi_\mathrm{dis}^\mathcal{S}[\langle U \rangle] 
= \textit{Constant}(T) > 0 \, .
\label{eq:chi_dis_finite}
\end{equation}

\textit{Effective packing fraction.} Now, we show that the disorder
fluctuations in the potential energy $\langle U \rangle (\sigma)$
and the empirical limit value for $\chi_\mathrm{dis}^\mathcal{S}[\langle
U \rangle]$, as given by Eq.~(\ref{eq:chi_dis_finite}), can be
explained by fluctuations in a single scalar variable, namely an
effective packing fraction $\phi_\mathrm{eff}$. The additive packing fraction $\phi_{\rm hs}$, cf.~Eq.~(\ref{eq_phihs}), is not an
appropriate measure of a packing fraction for the non-additive
soft-sphere system that we consider in this study. Therefore, we
define an effective packing fraction $\phi_\mathrm{eff}$ to take
into account the non-additivity of our model system.

The idea is to assign to each particle $i$ an ``average'' volume
$V_i$ that accounts for the non-additive interactions. For this
purpose, we first identify all $|\mathcal{N}_i|$ neighbors of $i$
within a given cutoff $r_c$,
\begin{equation}
\mathcal{N}_i = 
\left\{ ~j \in \{1,\dots,N\}~| ~j\neq i,~ r_{ij} < r_c \right\}\,.
\end{equation}
Here $r_c = 1.485$ is chosen, which corresponds to the location of
the first minimum of the radial distribution function at the
temperature $T=0.3$.  Then, the volume $V_i$ of particle $i$ is
defined as
\begin{equation}
V_i = \frac{1}{|\mathcal{N}_i|} 
\sum_{j \in \mathcal{N}_i} \frac{\pi}{6} \sigma_{ij}^3 \,,
\end{equation}
where non-additive diameters $\sigma_{ij}$ are given by
Eq.~(\ref{eq:non_additivity}).

Now we define an effective packing fraction $\phi_{\rm eff}$  as 
\begin{equation}
\phi_\mathrm{eff} = V^{-1}
\sum_{i=1}^{N} V_i \, .
\label{eq_phieff}
\end{equation}
Note that different from the hard-sphere packing fraction $\phi_{\rm
hs}$, the value of the effective packing fraction $\phi_{\rm eff}$
of a given sample not only depends on the diameters $\sigma_i$, but
it also depends on the coordinates ${\bf r}_i$. Thus, in our
simulations of glassforming liquids, it is a thermally fluctuating
variable. Therefore, we will use its thermal average $\langle
\phi_\mathrm{eff} \rangle$ in our analysis below.

An alternative effective packing fraction can be defined by assigning
an average diameter $S_i = \frac{1}{|\mathcal{N}_i|} \sum_{j \in
\mathcal{N}_i} \sigma_{ij}$ instead of an average volume $V_i$ to
each particle. The corresponding packing fraction is given by
\begin{equation}
\tilde{\phi}_\mathrm{eff} = V^{-1} 
\sum_{i=1}^{N} \frac{\pi}{6} S_i^3 \, .
\end{equation}
Below, we use the effective packing fractions $\phi_{\rm eff}$ and
$\tilde{\phi}_{\rm eff}$ to analyse the sample-to-sample fluctuations
in model $\mathcal{S}$. Although both definitions lead to similar
results, we shall see that $\phi_{\rm eff}$ seems to provide a
slightly better characterization of the thermodynamic state of the
system than $\tilde{\phi}_{\rm eff}$.

\begin{figure}
\includegraphics{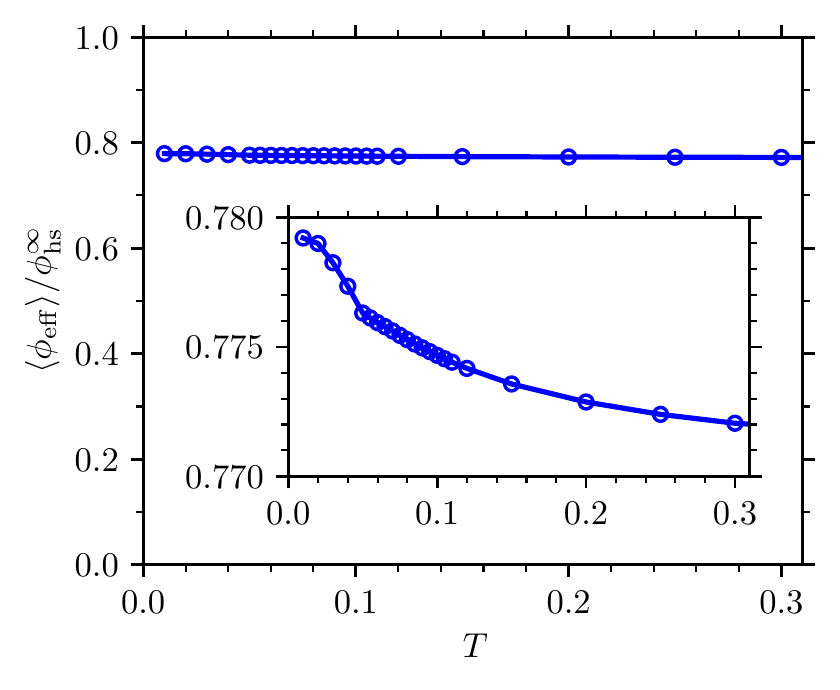}
\caption{Reduced effective packing fraction $\langle \phi_\mathrm{eff}
\rangle/\phi_{\rm hs}^\infty$ as a function of temperature $T$. The
inset zooms into a region around $\langle \phi_\mathrm{eff}
\rangle/\phi_{\rm hs}^\infty = 0.775$. \label{fig5}}
\end{figure}
Figure \ref{fig5} displays the temperature dependence of $\langle
\phi_\mathrm{eff} \rangle$. It is almost constant over the whole
considered temperature range. This is a plausible result when one
considers the weak temperature dependence of the structure of
glassforming liquids.  As we can infer from the inset of this figure,
$\langle \phi_\mathrm{eff} \rangle$ increases mildly from about
$0.772$ at $T=0.3$ to about $0.779$ at $T=0.01$.

\begin{figure}
\centering 
\includegraphics[width=\columnwidth]{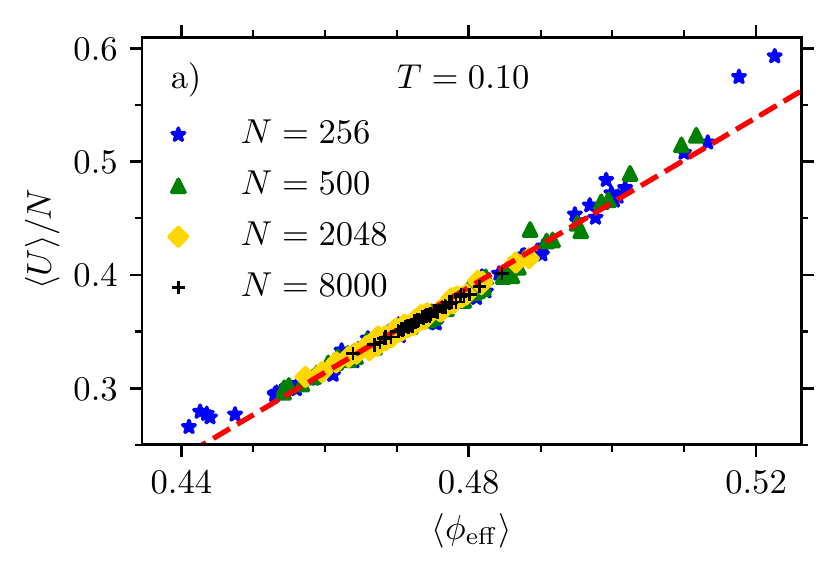}
\includegraphics[width=\columnwidth]{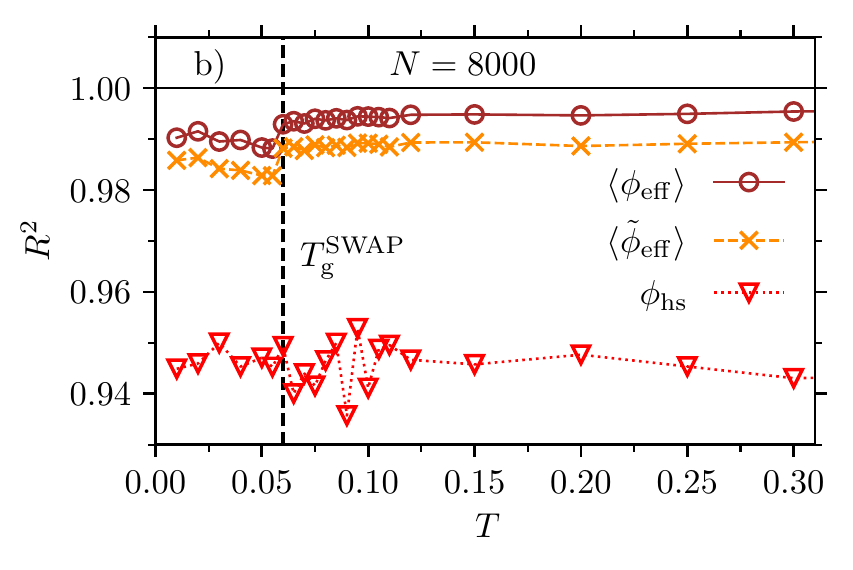}
\includegraphics[width=\columnwidth]{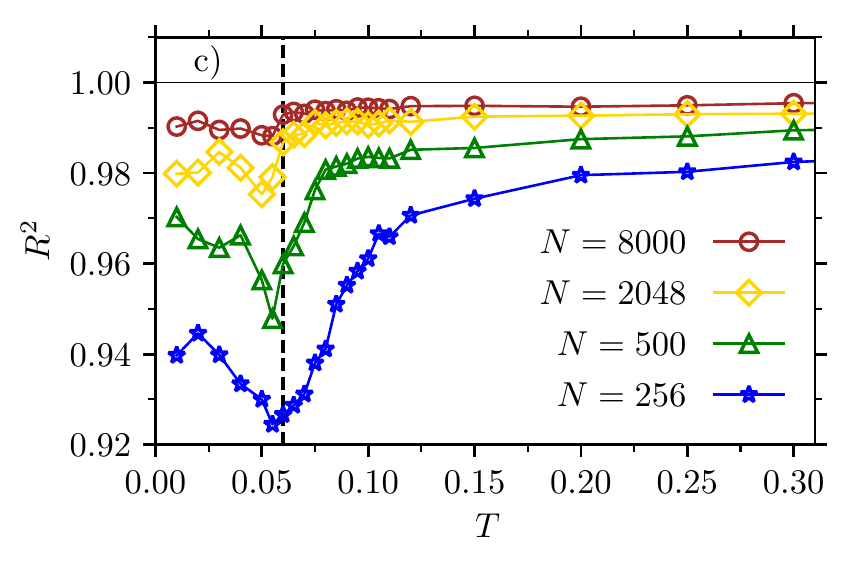}
\caption{a) Scatter plot showing data points $\left(\langle \phi_{\rm
eff} \rangle (\sigma)\,,\,\langle U \rangle(\sigma)/N \right)$ for
model $\mathcal{S}$ at $T=0.10$ and different system sizes $N$.
Each tuple belongs to a particular diameter realization $\sigma$.
The red line is obtained via a linear-regression model $\phi \to
\langle U \rangle$ with dependent variable $\langle U \rangle$ and
regressor $\phi = \langle \phi_{\rm eff} \rangle$ for $N=2048$.
Its coefficient of determination is $R^2 \approx 0.984$.  b)
Coefficient of determination $R^2$ of the linear regression model
$\phi \to \langle U \rangle$ as a function of $T$ for $N=8000$,
using $\phi = \phi_\mathrm{hs}$ (red triangles), $\langle \phi_{\rm
eff} \rangle$ (brown circles), and $\langle \tilde{\phi}_\mathrm{eff}
\rangle$ (orange crosses) as regressors $\phi$. c) Similar to b),
but here $R^2$ as a function of $T$ is shown for regressor $\phi =
\langle \phi_{\rm eff} \rangle$ only, however for different system
sizes $N$. \label{fig6}}
\end{figure}
Now, we will use the variable $\langle \phi_\mathrm{eff} \rangle$
to quantify the sample-to-sample fluctuations of the potential
energy per particle $\langle U \rangle(\sigma)/N$.

In Fig.~\ref{fig6}a, we show $\langle U \rangle(\sigma)/N$ as a
function of the mean packing fraction $\langle \phi_{\rm
eff}\rangle(\sigma)$ at the temperature $T = 0.10$.  Here, we have
used the data for $N=256$, $500$, and $2048$ particles.  The plot
suggests that the fluctuations of $\langle U \rangle$ can be explained
by the variation of $\langle \phi_{\rm eff}\rangle$.  We elaborate
this finding by calculating the coefficient of determination $R^2$
of a linear-regression fit with dependent variable $\langle U
\rangle/N$ and regressor $\langle \phi_{\rm eff}\rangle$.

In Fig.~\ref{fig6}b we show $R^2$ as a function of $T$ for the
system size $N=8000$. The linear regression analysis shows that
approximately $99.5\%$ of the fluctuations can be explained by
$\langle \phi_{\rm eff}\rangle$. This is a striking but physically
plausible result, as it shows how a reduction from $N$ degrees of
freedom given by $\sigma$ to one degree of freedom given by a
thermodynamically relevant parameter $\langle \phi_{\rm eff}\rangle$
is sufficient to explain nearly all of the fluctuations. Also
included in Fig.~\ref{fig6}b is the coefficient of determination
$R^2$ using $\phi = \phi_\mathrm{hs}$ and $\langle
\tilde{\phi}_\mathrm{eff} \rangle$ as a regressor. While we obtain
$R^2\approx 0.95$ for $\phi = \phi_\mathrm{hs}$, i.e.~clearly below
the value for $\langle \phi_{\rm eff}\rangle$, the value of $R^2$
for $\langle \tilde{\phi}_\mathrm{eff} \rangle$ is only slightly
smaller, $R^2 \approx 0.99$. Thus, among the three measures of the
packing fraction, the variable $\langle \phi_{\rm eff}\rangle$ gives
the best results. Note that the glass transition at $T_{\rm g}^{\rm
SWAP} \approx 0.06$ is associated with a small drop of $R^2$ for
the effective packing fractions.

Figure \ref{fig6}c displays the temperature dependence of $R^2$ for
$\langle \phi_{\rm eff}\rangle$ for different system sizes $N$.
The plot indicates a significant decrease of $R^2$ with decreasing
$N$, especially at low temperatures around the glass-transition
temperature $T_{\rm g}^{\rm SWAP} \approx 0.06$. The reason is that
a linear relationship between $\langle U \rangle(\sigma)/N$ and
$\langle \phi_{\rm eff}\rangle$ is expected to only hold in the
vicinity of the disorder-averaged value $\overline{\langle
\phi_\mathrm{eff} \rangle}$.  For small system sizes, however,
relatively large nonlinear deviations from this value occur that
are reflected in a lower value of the coefficient of determination
$R^2$. Moreover, for small $N$, the discretized nature of the
diameter configuration does not any longer allow a description in
terms of a single variable such as $\langle \phi_{\rm eff}\rangle$.

Our empirical results justify the idea to replace the dependency
of $\langle U \rangle$ on the diameter configuration $\sigma$ by
one on the single parameter $\langle \phi_{\rm eff}\rangle$,
\begin{align}
\langle U \rangle (\sigma) 
&\approx 
U^* \left(\langle \phi_\mathrm{eff} \rangle (\sigma) \right) 
\nonumber \\
&\approx U^*\left( \overline{\langle \phi_\mathrm{eff} \rangle} \right) 
+ \frac{\partial U^* }{\partial \phi}
\big|_{\phi= \overline{\langle \phi_\mathrm{eff} \rangle}} 
( \langle \phi_\mathrm{eff} \rangle - \overline{\langle 
\phi_\mathrm{eff} \rangle} ).
\label{eq_taylor_expansion_U}
\end{align}
Here $U^*$ is an unknown function in a scalar variable. According
to the Taylor expansion above, fluctuations in $\langle U \rangle$
are inherited from those in $\langle \phi_\mathrm{eff} \rangle$ as
\begin{align}
\mathrm{Var}( U^*) \approx
\left(\frac{\partial U^* }{\partial \phi}\right)^2 
\big|_{\phi= \overline{\langle \phi_\mathrm{eff} \rangle}} 
\mathrm{Var}(\langle \phi_\mathrm{eff} \rangle )\, .
\label{eq_variance_U}
\end{align}
Since $\langle \phi_\mathrm{eff} \rangle$ should scale similarly
to the additive hard-sphere packing fraction $\phi_{\rm hs}$, we
have $\mathrm{Var}(\langle \phi_\mathrm{eff} \rangle ) \propto 1/N$.
Then, since $U^*$ is extensive, Eq.~(\ref{eq:chi_dis_finite}) is
confirmed.

\section{Structural Relaxation}
\label{sec:structural_dynamics}
In this section, the dynamic properties of the models $\mathcal{S}$
and $\mathcal{D}$ are compared. To this end, we analyze a time-dependent
overlap function that measures the structural relaxation of the
particles on a microscopic length scale.  The timescale on which
this function decays varies from sample to sample; these fluctuations
around the average dynamics can be quantified in terms of a dynamic
susceptibility. We shall see that the susceptibility in model
$\mathcal{S}$ can be split into two terms. While the first term is
due to thermal fluctuations and also present in model $\mathcal{D}$,
the second term is due to the disorder in $\sigma$. At low temperatures,
the contribution from the disorder can be the dominant term in the
susceptibility.

For our analysis, we consider MD simulations in the microcanonical
ensemble as well as hybrid simulations, combining MD with the Swap
Monte Carlo technique (see Sec.~\ref{sec:simulation_details}). In
the following, we refer to these dynamics as ``$NVE$'' and ``SWAP'',
respectively.

\begin{figure*}
\includegraphics{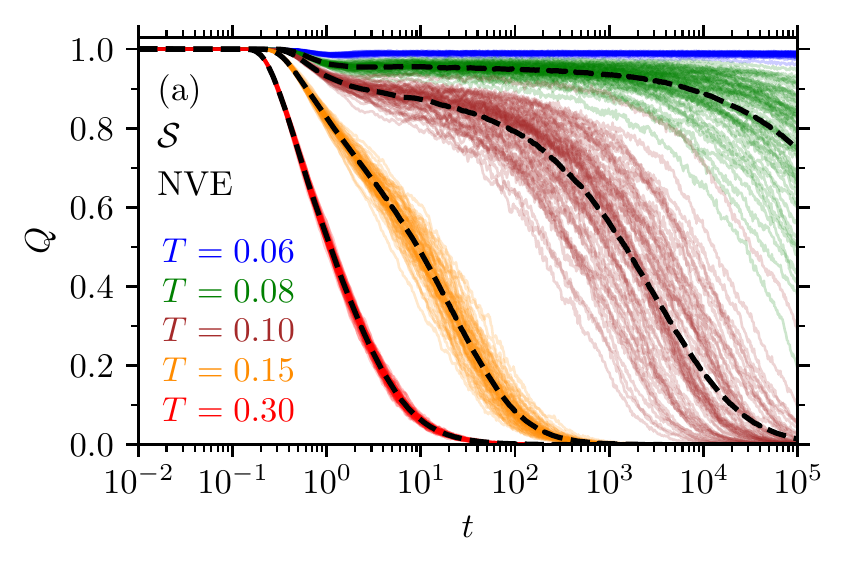}
\includegraphics{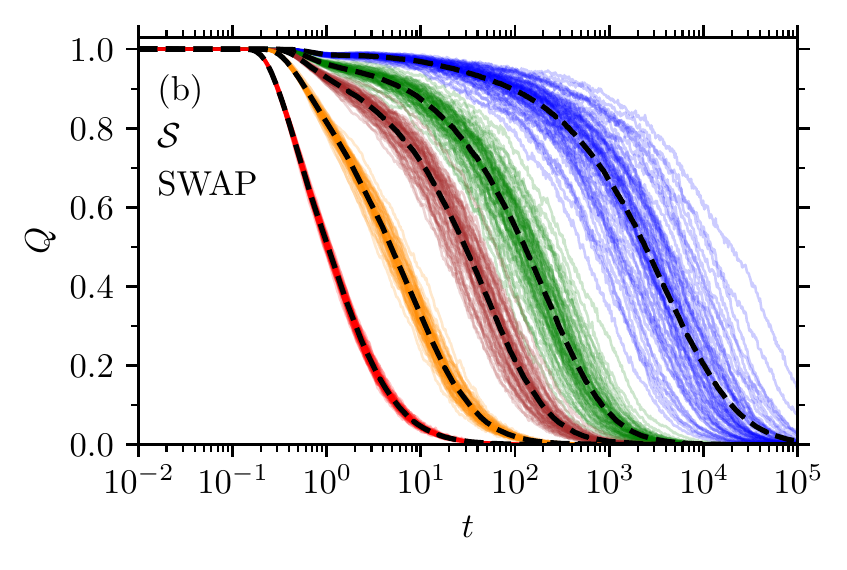}\\
\includegraphics{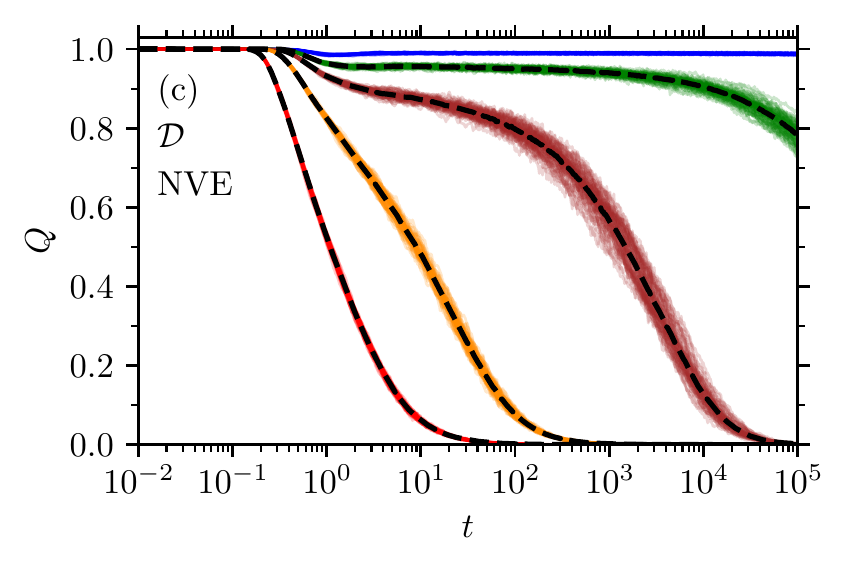}
\includegraphics{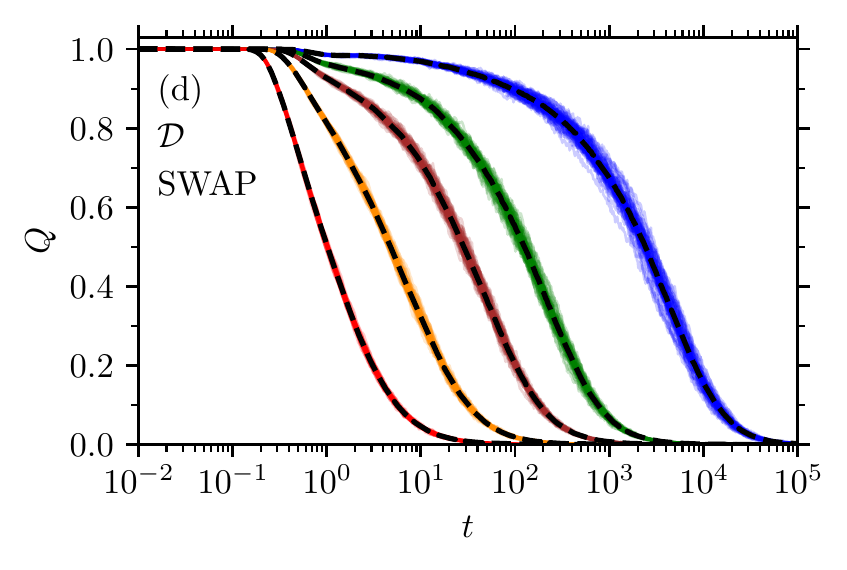}\\
\caption{Overlap $Q(t)$ as a function of time $t$ for $NVE$ (left
column) and SWAP dynamics (right column) for models $\mathcal{S}$
and $\mathcal{D}$. For the selected temperatures $T$ the initial
configurations are in equilibrium. Solid colored lines represent
$60$ individual simulations, while black dashed lines indicate their
sample average. All results correspond to systems with $N = 8000$
particles. \label{fig7}}
\end{figure*}
{\it Glassy dynamics.} A peculiar feature of the structural relaxation
of glassforming liquids is the cage effect. On intermediate timescales,
each particle gets trapped in a cage that is formed by its neighboring
particles.  To analyze structural relaxation from the cages, we
therefore have to look at density fluctuations on a length scale
$a$ similar to the size of the fluctuations of a particle inside
such a cage. On a single-particle level, a simple time-dependent
correlation function that measures the relaxation is the self part
of the overlap function, defined by
\begin{align}
Q(t) =  \frac{1}{N} \sum_{i=1}^{N} 
\Theta( a - |{\bf r}_i(t)- {\bf r}_i(0)|  ) \, .
\label{eq:Q_definition}
\end{align}
Here, we choose $a=0.3$ for the microscopic length scale. The
behavior of $Q(t)$ is similar to that of the incoherent intermediate
scattering function at a wave-number corresponding to the location
of the first-sharp diffraction peak in the static structure factor.
We note that we have not introduced any averaging in the definition
(\ref{eq:Q_definition}).  In the following, we will display the
decay of $Q(t)$ for 60 individual samples at different temperatures.
The corresponding initial configurations at $t=0$ were fully
equilibrated with the aid of the SWAP dynamics before, as explained
in Sec.~\ref{sec:simulation_details}.

Figure \ref{fig7} shows the overlap function $Q(t)$ for model
$\mathcal{S}$ and model $\mathcal{D}$, in both cases for the $NVE$
and the SWAP dynamics.  In all cases, we can see the typical
signatures of glassy dynamics.  At a high temperature, $T=0.3$, the
function $Q(t)$ exhibits a monotonous decay to zero on a short
microscopic timescale.  Upon decreasing the temperature first a
shoulder and then a plateau-like region emerges on intermediate
timescales. This plateau extends over an increasing timescale with
decreasing temperature and indicates the cage effect. Particles are
essentially trapped within the same microstate in which they were
initially at $t=0$. At the high temperature $T=0.3$ the decay of
$Q(t)$ is very similar for $NVE$ and SWAP dynamics. Towards low
temperatures, however, the decay is much faster in the case of the
SWAP dynamics, as expected. A striking result is that at lower
temperatures, the individual curves in model $\mathcal{S}$ show
much larger variation than those in model $\mathcal{D}$. In the
following, these sample-to-sample fluctuations shall be quantified
in terms of a dynamic susceptibility.

\begin{figure}
\includegraphics{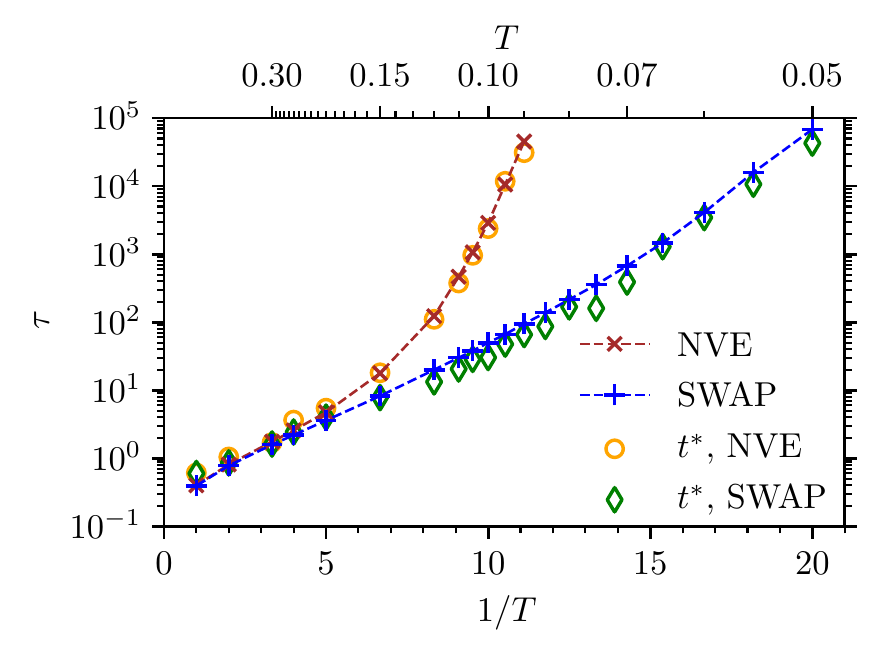}%
\caption{Relaxation time $\tau$, as extracted from the expectation
of the overlap function, $\mathrm{E}[Q](t)$, and the time $t^* =
\arg \max_{t} \chi(t)$, where the maximum of the dynamic susceptibility
$\chi(t)$ occurs, for $NVE$ and SWAP dynamics. Here, a system with
$N=8000$ particles is considered.  \label{fig8}}
\end{figure}

\textit{Relaxation time $\tau$.} From the expectation of the overlap
function, $\mathrm{E}[Q](t)$ (black dashed lines in Fig.~\ref{fig7}),
we extract an alpha-relaxation time $\tau$, defined by $\mathrm{E}[Q](\tau)
= 1/{\rm e}$. In Fig.~\ref{fig8}, the logarithm of the timescale
$\tau$ as a function of inverse  temperature $1/T$ is shown. Also
included in this plot are the times $t^*$ where the fluctuations
of $Q(t)$ are maximal, which will be discussed in the following
paragraph ``Dynamic susceptibility''. One observes an increase of
$\tau$ by about five orders of magnitude upon decreasing $T$. This
increase is much quicker for the $NVE$ than for the SWAP dynamics,
reflecting the fact that $T_{\rm g}^{\rm SWAP}$ is much lower than
$T_{\rm g}^{NVE}$ (cf.~Fig.~\ref{fig3}). The glass-transition
temperatures defined in Sec.~\ref{sec:thermodynamics} via the drop
in the specific heat $C_V(T)$ are approximately consistent with the
alternative definition via $\tau(T_{\rm g}) = 10^5$.

\begin{figure*}
\includegraphics{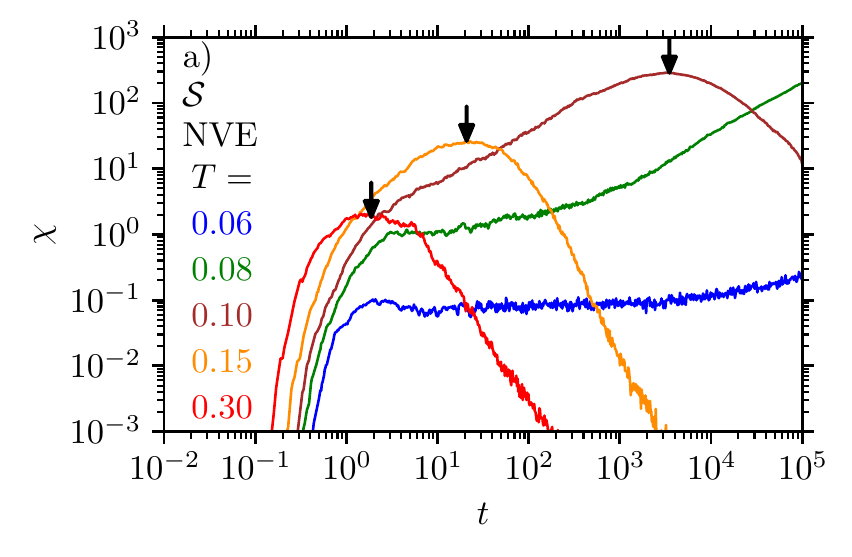}
\includegraphics{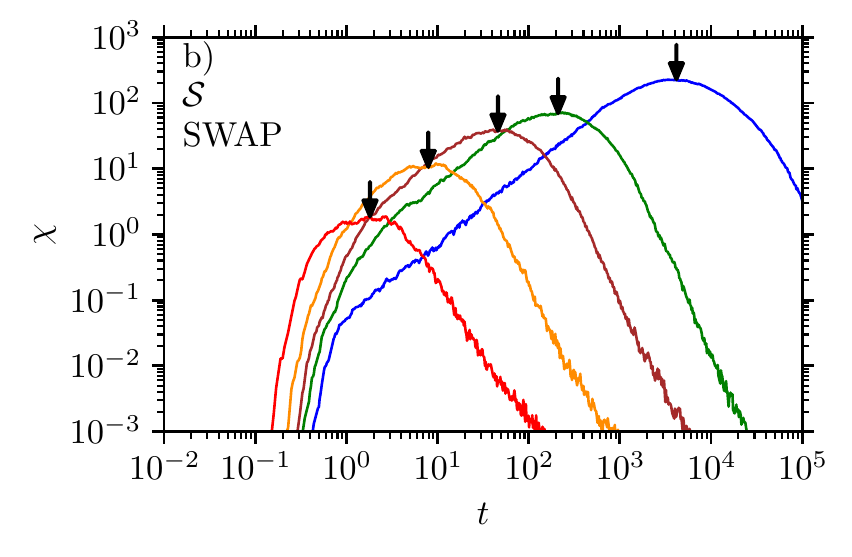}\\
\includegraphics{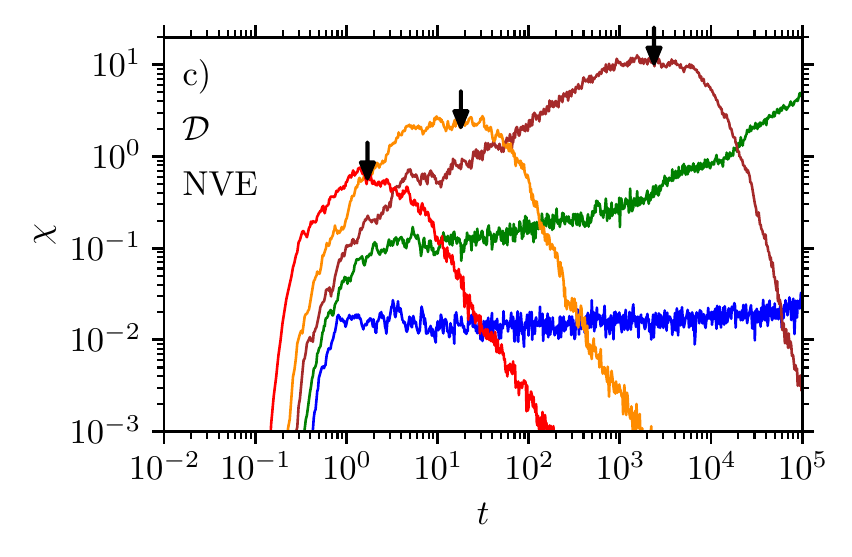}
\includegraphics{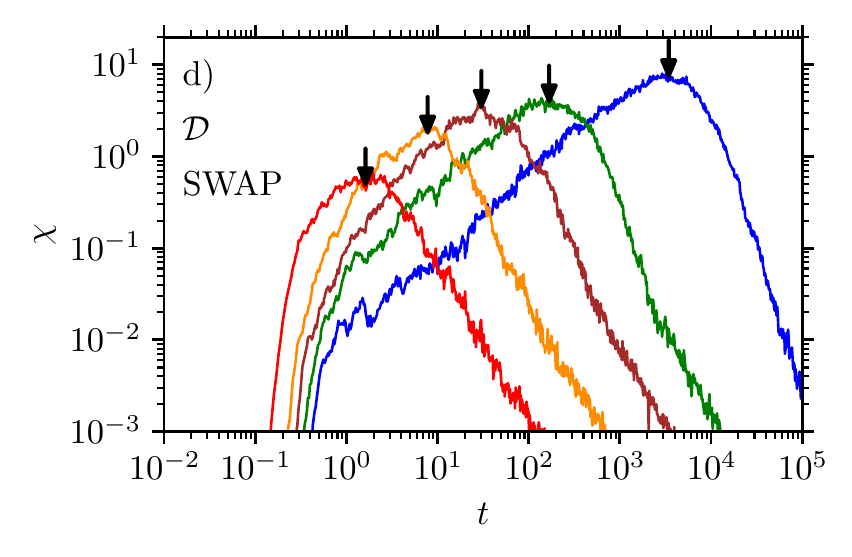}\\
\caption{Dynamic susceptibility $\chi$ as a function of time $t$
for different temperatures $T$ and systems with $N=8000$ particles.
Results for all four combinations of $NVE$ and SWAP dynamics with
models $\mathcal{S}$ and $\mathcal{D}$ are shown, as labeled in
a)-d).  Maxima of $\chi(t)$ are marked by arrows. Prior to their
calculation we performed a moving average over the raw data.
\label{fig9}}
\end{figure*}
\newpage
{\textit{Dynamic susceptibility $\chi(t)$.}} A characteristic feature
of glassy dynamics is the presence of dynamical heterogeneities
that are associated with large fluctuations around the ``average''
dynamics.  These fluctuations can be quantified in terms of a dynamic
(or four-point) susceptibility. For the overlap function $Q(t)$,
this susceptibility $\chi(t)$ can be defined as
\begin{align}
\chi(t) = 
N \mathrm{Var}\left(Q(t)\right) \, .
\label{Eq:dynamic_sus_Q}
\end{align}
The function $\chi(t)$ measures the fluctuations of $Q(t)$ around
the average $\mathrm{E}[Q](t)$. In practice, we use the data of
$Q(t)$ from the ensemble of 60 independent samples.

Figure \ref{fig9} shows the dynamic susceptibility $\chi(t)$ for
the same cases as for $Q(t)$ in Fig.~\ref{fig7}. As a common feature
of glassy dynamics~\cite{chandler2006, cavagna2009}, $\chi(t)$
exhibits a peak $\chi^* := \mathrm{max}_t~\chi(t)$ at $t = t^*$.
The timescale $t^*$ is roughly equal to the alpha-relaxation time
$\tau$, see Fig.~\ref{fig8}. At the temperatures $T=0.1$ for the
$NVE$ and $T=0.06$ for the SWAP dynamics, $\chi^*$ is more than one
order of magnitude larger for model $\mathcal{S}$ than for model
$\mathcal{D}$. This indicates that the disorder in $\sigma$ of model
$\mathcal{S}$ strongly affects the sample-to-sample fluctuations.
In the following paragraph ``Variance decomposition'' we will present
how one can distinguish disorder from thermal fluctuations.

\begin{figure}
\includegraphics{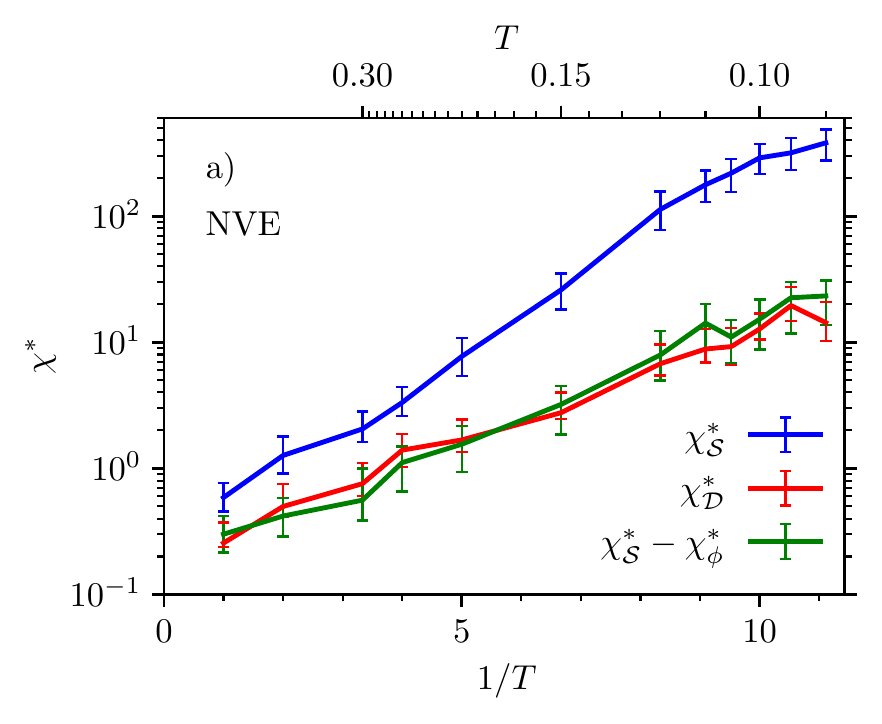}
\includegraphics{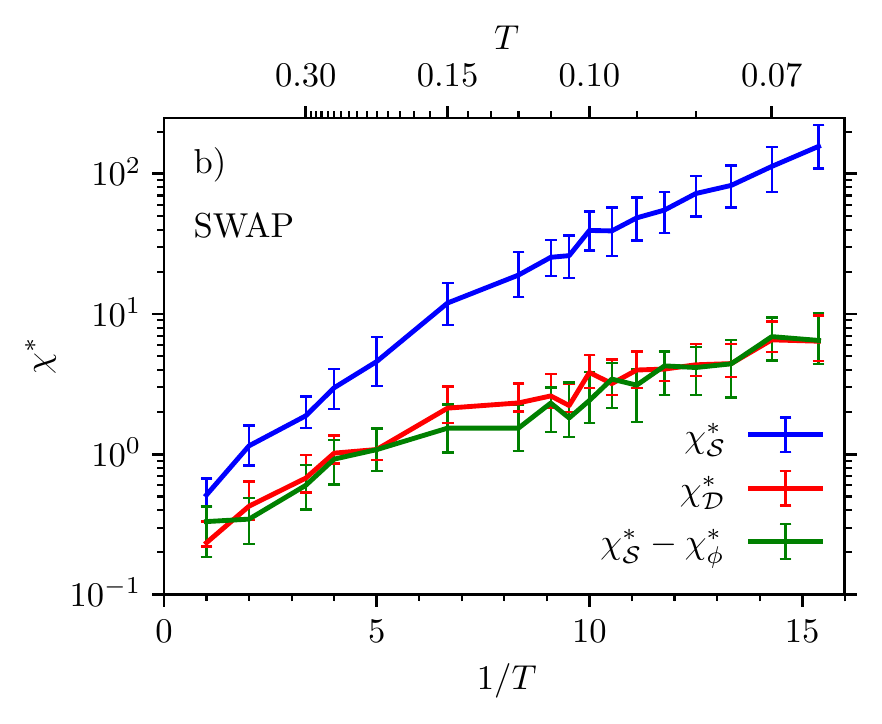}
\caption{Maximum of the dynamic susceptibility, $\chi^* = \mathrm{max}_t
\chi (t)$, as a function of $1/T$ for a) the $NVE$ and b) the SWAP
dynamics. Results are shown for models $\mathcal{S}$ (blue line)
and $\mathcal{D}$ (red line) with $N=8000$ particles. The green
solid line displays $\chi_\mathcal{S}^* - \chi_\phi^*$, i.e.~the
total susceptibility minus the explained part caused by the
packing-fraction fluctuations. \label{fig10}}
\end{figure}
Figure~\ref{fig10} shows the maximum of the dynamic susceptibility,
$\chi^*$, as a function of inverse temperature, $1/T$, for $NVE$
and SWAP dynamics. In both cases, the results for model $\mathcal{S}$
($\chi_\mathcal{S}^*$) and model $\mathcal{D}$ ($\chi_\mathcal{D}^*$)
are included, considering systems with $N=8000$ particles. In all
cases $\chi^*$ increases with decreasing temperature $T$, as expected
for glassy dynamics. For both types of dynamics the difference
$\Delta \chi^* = \chi_\mathcal{S}^* - \chi_\mathcal{D}^*$ increases
with decreasing temperature as well. The lowest temperatures for
which we can calculate $\Delta \chi^*$ are (i) $T= 0.09$ with a
relative deviation $\Delta \chi^*/\chi^*_\mathcal{D} \approx 18$
for the $NVE$ and (ii) $T = 0.065$ with $\Delta \chi^*/\chi^*_\mathcal{D}
\approx 23$ for the SWAP dynamics.

{\textit{Variance decomposition}.} To understand the difference
$\Delta \chi^*$ between $\chi_\mathcal{S}$ and $\chi_\mathcal{D}$,
we will decompose the dynamic susceptibility $\chi_\mathcal{S}$ of
model $\mathcal{S}$ into one term that stems from the thermal
fluctuations of the phase-space variables, and a second term that
is caused by the sample-to-sample variation of the diameters $\sigma$.

As a matter of fact, in model $\mathcal{S}$ the overlap function
$Q(t)$ and similar correlation functions depend on \textit{two
random vectors}, namely the initial phase-space point $q_0 =
\left(r(0),v(0)\right)$ \textit{and} the diameters $\sigma$. As a
consequence, we define and calculate $\chi = N \mathrm{Var}(Q)$ on
a probability space with respect to the joint-probability density
\begin{align}
\rho( q_0, \sigma) = \rho(q_0|\sigma)g(\sigma).
\label{eq_joint_probability}
\end{align}
Here $\rho(q_0|\sigma)$ is the conditional phase-space density
introduced in Eq.~(\ref{eq_condpsd}) and $g(\sigma)$ is the diameter
distribution defined by Eq.~(\ref{eq:density_sigma_global}).

Now, since $Q$ depends on two random vectors $q_0$ and $\sigma$,
we can decompose $\chi = N \mathrm{Var}(Q)$ according to the
\textit{variance decomposition formula}, also called \textit{law
of total variance} or \textit{Eve's law} \cite{chung1974}:
\begin{align}
\mathrm{Var}(Q) &=
\mathrm{E}\left[ \mathrm{Var}(Q|\sigma) \right] +
\mathrm{Var}\left( \mathrm{E}[Q|\sigma] \right) \label{eq_eve1}\\
&\equiv \overline{ \langle Q^2 - \langle Q \rangle^2 \rangle } +
\overline{ \langle Q \rangle^2 - \overline{ \langle Q \rangle}^2} \, .
\label{eq_eve2}
\end{align}
Here, $\mathrm{E}\left[ \mathrm{Var}(Q|\sigma) \right]$ describes
intrinsic thermal fluctuations, while the term $\mathrm{Var}\left(
\mathrm{E}[Q|\sigma] \right)$ expresses fluctuations induced by the
disorder in $\sigma$.

The first summand in Eq.~(\ref{eq_eve1}) is expected to coincide
for both models $\mathcal{S}$ and $\mathcal{D}$ for sufficiently
large $N$, as $\mathrm{Var}(Q|\sigma)$ describes intrinsic thermal
fluctuations for a given realization of $\sigma$, which are calculated
via the \textit{model-independent} conditional phase-space density
$\rho(q_0|\sigma)$. The physical observable $\mathrm{Var}(Q|\sigma)$
should not depend on microscopic details of the diameter configuration
$\sigma$ for sufficiently large $N$. For the cumulative distribution
functions of the diameters, the consistency equation $\lim_{N\to\infty}
F_N^\mathcal{S}(s) = F(s) = \lim_{N\to\infty} F_N^\mathcal{D}(s)$
holds. Thus, we expect that $\mathrm{E}^\mathcal{S}\left[
\mathrm{Var}(Q|\sigma) \right] \approx \mathrm{E}^\mathcal{D}\left[
\mathrm{Var}(Q|\sigma) \right]$.  This equation should be exact in
the limit $N\to\infty$. We have implicitly used this line of argument
also in Sec.~\ref{sec:thermodynamics}, where we have only shown
numerical results of the specific heat for model $\mathcal{D}$.
Furthermore, for model $\mathcal{D}$ we have exactly
$\mathrm{E}^\mathcal{D}[\mathrm{Var}(Q|\sigma)] =
\mathrm{Var}(Q|\sigma^\mathcal{D}) = \mathrm{Var}^\mathcal{D}(Q)$,
since here there is only one diameter configuration $\sigma =
\sigma^\mathcal{D}$ for a given system size $N$.

Summarizing the results above, we can express the dynamic susceptibility
for model $\mathcal{S}$ as follows:
\begin{equation}
\mathrm{Var}^\mathcal{S}(Q) = \mathrm{Var}^\mathcal{D}(Q) +
\mathrm{Var}^\mathcal{S}\left( \mathrm{E}[Q|\sigma] \right) 
\label{eq_VarS_VarD_Dis}. 
\end{equation}
Now the aim is to estimate the second summand in
Eq.~(\ref{eq_VarS_VarD_Dis}).  We assume that we can describe the
disorder in $\sigma$ by a single parameter, namely the thermally averaged 
effective packing fraction $\langle \phi_{\rm eff}
\rangle(\sigma)$, defined by Eq.~(\ref{eq_phieff}).  This idea has
been already proven successful in Sec.~\ref{sec:thermodynamics},
when we described the disorder fluctuations of the potential energy.
Similarly, we write
\begin{align}
\mathrm{E}[Q|{\sigma}]
\equiv
\langle Q \rangle (\sigma)
\approx
Q^*({\langle \phi_{\rm eff} \rangle}(\sigma))
\, ,
\label{eq_var2}
\end{align}
assuming that the values of $\langle Q \rangle(\sigma)$, which
depend on $N$ degrees of freedom, can be described by a function
$Q^*$ that only depends on a scalar argument, the scalar-valued
function $\langle \phi_{\rm eff} \rangle(\sigma)$. The function
$Q^*$ is unknown, but can be estimated numerically  with a
linear-regression analysis, predicting $\langle Q \rangle$ with the
regressor $\langle \phi_{\rm eff}\rangle$.  Insertion of
Eq.~(\ref{eq_var2}) into Eq.~(\ref{eq_VarS_VarD_Dis}) gives
\begin{align}
\mathrm{Var}^\mathcal{S}(Q)
\approx
\mathrm{Var}^\mathcal{D}(Q)  + 
\mathrm{Var}^\mathcal{S}( Q^*(\langle \phi_{\rm eff} \rangle) ) \, .
\label{eq:variance_decomp_approx1}
\end{align}
We can write this equation in terms of susceptibilities,
\begin{align}
\chi_\mathcal{S}
&~\approx
\chi_\mathcal{D}  + \chi_\mathcal{\phi},\\
\chi_\mathcal{\phi} 
&:= N\mathrm{Var}^\mathcal{S}( Q^*(\langle \phi_{\rm eff} \rangle) ).
\end{align}
Along the lines of Eq.~(\ref{eq_variance_U}) in
Sec.~\ref{sec:thermodynamics}, we can expand the overlap function
$Q^*$ around $\overline{\langle \phi_{\rm eff} \rangle}$ to obtain
\begin{align}
\mathrm{Var}^\mathcal{S}( Q^*(\langle \phi_{\rm eff} \rangle)
\approx \mathrm{Var}^\mathcal{S}(\langle \phi_{\rm eff} \rangle) \,
\left(\frac{\partial Q^*(\phi ) }{\partial  \phi }
\big|_{\phi = \overline{\langle \phi_\mathrm{eff} \rangle}} \right)^2.
\label{eq:VarQ_Varphi_inheritence}
\end{align}
Since $\mathrm{Var}^\mathcal{S}( \langle \phi_{\rm eff} \rangle  )
\sim \mathrm{Var}^\mathcal{S}(  \phi_{\rm hs}  ) \propto N^{-1}$
and $Q^* \sim Q \in \mathcal{O}(1)$, this equation implies that the
susceptibility $\chi_\phi$, to leading order, does not depend on
$N$.  Moreover, for a given temperature $T$ and time $t$, it
approaches a constant value in the thermodynamic limit, $N\to
\infty$. For small system sizes, however, higher-order corrections
to Eq.~(\ref{eq:VarQ_Varphi_inheritence}) cannot be neglected.
Beyond that, the discretized nature of the system at small $N$ will
lead to a failure of the ``continuity-assumption'' (\ref{eq_var2})
itself.  Finite-size effects of $\chi$ will be analyzed in the
following paragraph.

In Fig.~\ref{fig10}, we show for the system with $N=8000$ particles
that $\chi_\phi^*$, i.e.~$\chi_\phi$ evaluated at $t = t^*$, indeed
captures the sample-to-sample fluctuations in model $\mathcal{S}$
due to the disorder in $\sigma$. Both for $NVE$ and SWAP dynamics,
it quantitatively describes the gap between $\chi_\mathcal{S}^*$
and $\chi_\mathcal{D}^*$.

\begin{figure}
\centering
\includegraphics{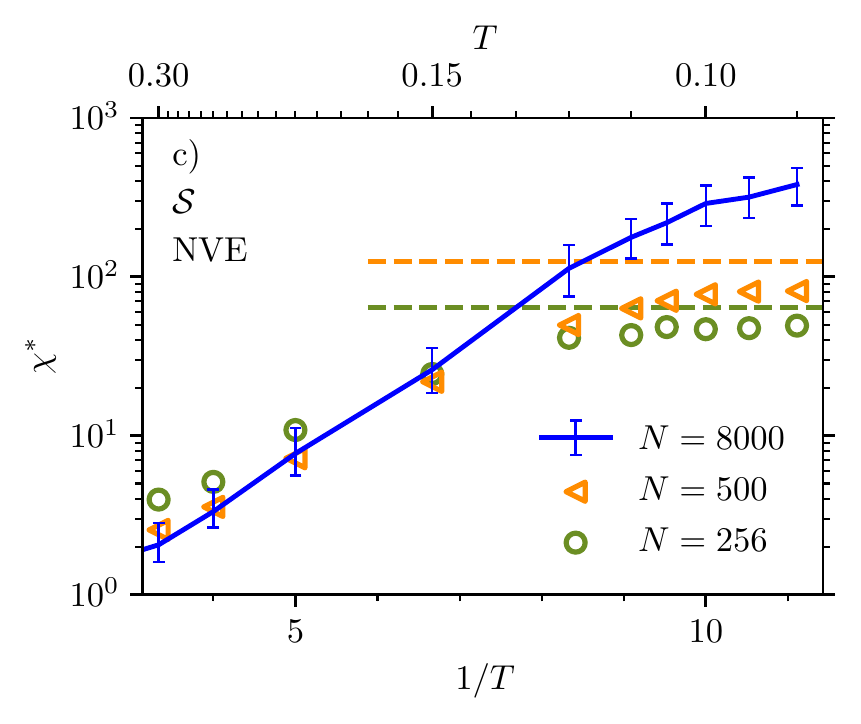}
\caption{$\chi_{\mathcal{S}}^*$ as a function of $1/T$ for different
system sizes $N$ using $NVE$ dynamics. The dashed lines denote
$N/4$, which is the upper bound according to \textit{Popoviciou's
inequality on variances}, see Eq.~(\ref{eq:Popoviciou}).  \label{fig11}}
\end{figure}

{\textit{Finite-size effects: Popoviciou's inequality on variances}.}
Here,  we analyze finite-size effects of the dynamic susceptibility
$\chi$.  To this end, we again consider the temperature dependence
of the maximum of the dynamic susceptibility, $\chi^*$, considering
only the case of the $NVE$ dynamics. Note that for model $\mathcal{D}$
finite-size effects in the considered temperature range $0.09 \le
T \le 0.3$ are negligible; therefore we only discuss model $\mathcal{S}$
in the following.

Figure~\ref{fig11} shows $\chi_{\mathcal{S}}^*$ as a function of
$1/T$ for $N=256$, 500, and 8000. At high temperatures $T$, where
fluctuations are small, there is hardly, if any, dependency on the
system size $N$. However, upon lowering $T$ a saturation occurs at
least for the small systems.  This behavior can be understood by a
\textit{hard} stochastic upper limit on fluctuations, which is given
by \textit{Popoviciou's inequality on
variances}~\cite{popoviciu1935equations}. This inequality is valid
for \textit{any bounded} real-valued random variable $X$: Let $c$
and $C$ be the lower and upper bound of $X$, respectively, then
Popoviciou states that $\mathrm{Var}(X) \leq (C^2 - c^2)/4$.  Applying
this result to $X=Q$ with sharp boundaries $c=0$ and $C=1$ yields
\begin{align}
\chi \equiv N \mathrm{Var}(Q) \leq N/4. 
\label{eq:Popoviciou}
\end{align}
Our data shows that this upper bound is quite sharp for $N=256$ and
$N=500$ at low $T$. This can be understood by the fact that the
\textit{equality} of the inequality~(\ref{eq:Popoviciou}) holds
precisely when $Q$ is a Bernoulli variable, i.e.~when there are
exactly two outcomes $Q=0$ or $Q=1$ each with probability $1/2$.
In this sense, the saturation of $\chi$ should occur at temperatures
$T$ and system sizes $N$ at a given $t$ when $Q(t)$ for approximately
half of the samples has decayed close to $0$ while for the other
half $Q$ is still close to $1$.

The inequality (\ref{eq:Popoviciou}) is very useful to estimate how
large a system size $N$ needs to be to avoid this kind of finite-size
effect: All one has to do is to compare the measured $\chi$ at a
given $N$ to the number $\chi_c := N/4$. In the case that $\chi
\approx \chi_c$, one has to consider larger system sizes $N$.

\section{Summary and conclusions}
\label{sec:conclusions}
In this work, we use molecular dynamics (MD) computer simulation
in combination with the SWAP Monte Carlo technique to study a
polydisperse model glassformer that has recently been introduced
by Ninarello {\it et al.}~\cite{ninarello2017}. Two methods are
used to choose the particle diameters $\sigma_1, \dots, \sigma_N$
to obtain samples with $N$ particles. Both of these approximate the
desired distribution density $f(\sigma)\sim \sigma^{-3}$ with their
histogram.  In model $\mathcal{S}$ the diameters are drawn from
$f(\sigma)$ in a stochastic manner. In model $\mathcal{D}$ the
diameters are obtained via a deterministic scheme that assigns an
appropriate set of $N$ values to them. We systematically compare
the properties of model $\mathcal{S}$ to those of model $\mathcal{D}$
and investigate how the sample-to-sample variation of the diameters
in model $\mathcal{S}$ affects various quantities: (i) classical
phase-space functions such as the potential energy $U$ and its
fluctuations, and (ii) dynamic correlation functions such as the
overlap function $Q(t)$ and its fluctuations as well.

Obviously, model $\mathcal{D}$ has the advantage that always ``the
most representative sample''~\cite{santen2001liquid} is used for
any system size $N$, while model $\mathcal{S}$ may suffer from
statistical outliers, especially in the case of small $N$. This
indicates that the quenched disorder introduced by the different
diameter configurations in model $\mathcal{S}$ may strongly affect
fluctuations that we investigate systematically in this work.

Our main findings can be summarized as follows: The sample-to-sample
fluctuations in model $\mathcal{S}$ can be described in terms of a
single scalar parameter, namely the effective packing fraction
$\langle \phi_{\rm eff} \rangle(\sigma)$, defined by Eq.~(\ref{eq_phieff}).
In terms of this parameter, one can explain the disorder fluctuations
of the potential energy (cf.~Fig.~\ref{fig6}) as well as the gap
between the dynamic susceptibilities of models $\mathcal{S}$ and
$\mathcal{D}$ (cf.~Fig.~\ref{fig10}). The sample-to-sample fluctuations
of the potential energy in model $\mathcal{S}$ can be quantified
in terms of the disorder susceptibility $\chi_\mathrm{dis}^\mathcal{S}$
which is a non-trivial function of temperature (cf.~Fig.~\ref{fig4})
and finite in the thermodynamic limit $N\to \infty$. In model
$\mathcal{S}$, at very low temperatures, the dynamic susceptibility
is dominated by the fluctuations due to the diameter disorder. Thus,
if one is aiming at analyzing the ``true'' dynamic heterogeneities
of a glassformer, that stem from the intrinsic thermal fluctuations,
one may preferentially use model $\mathcal{D}$. Note that it is
possible to calculate the same thermal susceptibility in model
$\mathcal{S}$ as in model $\mathcal{D}$, however the calculation
in $\mathcal{S}$ is more difficult, as it demands an additional
average over the disorder, as shown in Sec.~\ref{sec:structural_dynamics}.
This implies that model $\mathcal{S}$ requires more sampling in
this case.

Our findings are of particular importance regarding recent simulation
studies of polydisperse glassforming systems in external fields
\cite{guiselin2020overlap, RFIM_in_glassforming_liquid, lamp2022,
lerner2019, rainone2020} where a model $\mathcal{S}$ approach was
used to select the particle diameters. However, in these works
sample-to-sample fluctuations due to the disorder in $\sigma$ have
been widely ignored.  Exceptions are the studies by Lerner {\it et
al.}~\cite{lerner2019, rainone2020} where samples whose energy
deviates from the mean energy by more than 0.5\% were just discarded.
Here the use of a model $\mathcal{D}$ scheme would be a more efficient
alternative. However, one should still keep in mind that with regard
to a realistic description of experiments on polydisperse colloidal
systems, it might be more appropriate to choose model $\mathcal{S}$.

\appendix
\section{Convergence of the CDF $F_N^\mathcal{D}$
\label{app:Convergence_CDF_D}}
Here, we prove that the empirical cumulative distribution function
(CDF) $F_N^\mathcal{D}$ of model $\mathcal{D}$, see
Eqs.~(\ref{eq:CDF_empirical_sigma}) and
(\ref{eq:poly_dispersity_diameters_D}), converges uniformly to the
exact CDF $F$, defined by Eq.~(\ref{eq:CDF_sigma}). As we shall see
below, the order of convergence is at least $1$. For the strictly
monotonic function $\theta$, that we have introduced in
Sec.~\ref{sec:models}, we assume that it is strictly increasing,
but the proof is analogous for a strictly decreasing $\theta$.

In the first step, we show that
\begin{align}
\sigma_i \in [s_{i-1},s_i], && i=1,\dots,N.
\label{eq:sigma_model_D_estimate}	
\end{align}
Starting point is Eq.~(\ref{eq:poly_dispersity_diameters_D}) from 
which we estimate
\begin{align}
\theta(\sigma_i) 
&\leq N \int_{s_{i-1}}^{s_i} \theta(s_i) f(\sigma) ~\mathrm{d} \sigma\\
&= N \theta(s_i) \int_{s_{i-1}}^{s_i} f(\sigma)~ \mathrm{d} \sigma\\
&= N \theta(s_i) [F(s_{i})- F(s_{i-1})] \\ 
&= N \theta(s_i) \left[\frac{i}{N} - \frac{i-1}{N} \right] = \theta(s_i).
\end{align}
Since $\theta$ is strictly increasing, its inverse $\theta^{-1}$
exists and is strictly increasing, too. Applying $\theta^{-1}$ to
the inequality above yields $\sigma_i \leq s_i$. Similarly, we
obtain $\sigma_i \geq s_{i-1}$.  This confirms
Eq.~(\ref{eq:sigma_model_D_estimate}).

In the second step, we consider an arbitrary $\epsilon > 0$ and
natural numbers $N \geq N_0$ with $ N_0 = \lceil \epsilon^{-1}
\rceil$. Now we select $\sigma \in \mathbb{R}$. For $\sigma <
\sigma_{\rm m}$ or $\sigma > \sigma_{\rm M}$, we trivially have
$F_N^\mathcal{D}(\sigma) = F(\sigma)$.  In the remaining case
$\sigma_{\rm m} \leq \sigma \leq \sigma_{\rm M}$, an index $i$
exists such that $s_{i-1} \leq \sigma \leq s_i$.  The latter statement
is true, because the union of all intervals $[s_{i-1},s_{i}]$ yields
the total interval $[\sigma_{\rm m},\sigma_{\rm M}]$.  From
Eq.~(\ref{eq:sigma_model_D_estimate}) it follows that there are
exactly $i$ or $i-1$ particles with $\sigma_i \leq \sigma$, so that
$F_N^\mathcal{D}(\sigma) = i/N$ or $F_N^\mathcal{D}(\sigma) =
(i-1)/N$, respectively.

In the third step, we point out that $F(\sigma)$ is a monotonously
increasing function so that
\begin{align}
\frac{i-1}{N}  = F(s_{i-1}) \leq F(\sigma) \leq F(s_i) = \frac{i}{N}.
\end{align}
Subtracting $F_N^\mathcal{D}(\sigma)$ yields
\begin{align}
|F_N^\mathcal{D}(\sigma) - F(\sigma) | \leq 1/N \leq 1/N_0 
< \epsilon \,. \label{eq:F_N_D_convergence}
\end{align}
This proves the uniform convergence
\begin{align}
\lim_{N \to \infty} F_N^\mathcal{D} = F
\end{align}
of the order of convergence of at least $1$.

\section{Convergence of the CDF $F_N^\mathcal{S}$
\label{app:CDF_order_of_convergence}}
To find the order of convergence for $\lim_{N \to \infty} F_N^\mathcal{S}
= F$ of model $\mathcal{S}$, we measure deviations by $\Delta F =
(\mathrm{E}^\mathcal{S}[(F_N^\mathcal{S} - F)^2])^{1/2}$, see
Eq.~(\ref{eq:Delta_F}). We first calculate
\begin{align}
(F_N^\mathcal{S} - F)^2
&= 
\frac{1}{N^2} \sum_{i=1}^N \sum_{j=1}^N
(\mathbf{1}_i- F) (\mathbf{1}_j - F), \label{eq:CDF_N_S_convergence1}\\
\mathbf{1}_i(\sigma)
&:= 
\mathbf{1}_{ \left( -\infty, \sigma \right]}(\sigma_i). 
\end{align}
Here, we abbreviated the full notation of the indicator function 
$\mathbf{1}$. Its expectation is given by
\begin{align}
\mathrm{E}^\mathcal{S}[\mathbf{1}_i(\sigma)] 
&=
1\,P(\sigma_i \leq \sigma) + 0\,P(\sigma_i > \sigma) 
= F(\sigma). 
\label{eq:indicator_expectation}
\end{align}
Here, $P$ denotes the appropriate probability for model $\mathcal{S}$.
When calculating the expectation $\mathrm{E}^\mathcal{S}$ of
Eq.~(\ref{eq:CDF_N_S_convergence1}), only the diagonal terms $i=j$
remain due to the stochastic independence of the diameters $\sigma_i$
and $\sigma_j$ for $i\neq j$.  We end up with
\begin{align}
\mathrm{E}^\mathcal{S}[(F_N^\mathcal{S} - F)^2] 
&= F(1-F) N^{-1},\\
\Rightarrow \Delta F^\mathcal{S} &= \left((F(1-F)\right)^{1/2} \, N^{-1/2}.
\end{align}
This means the order of convergence for model $\mathcal{S}$ is only
$1/2$.  Concerning the prefactor, we have $\max_\sigma F(1-F) =
1/4$ at the $\sigma$ where $F(\sigma)=1/2$. Thus it is
\begin{align}
\max_\sigma \Delta F^\mathcal{S} = \frac{1}{2} N^{-1/2}.
\end{align}
Note that no inequality is used in the calculations above and thus the
order of convergence is sharp.


\begin{thebibliography}{35}
\expandafter\ifx\csname natexlab\endcsname\relax\def\natexlab#1{#1}\fi
\expandafter\ifx\csname bibnamefont\endcsname\relax
  \def\bibnamefont#1{#1}\fi
\expandafter\ifx\csname bibfnamefont\endcsname\relax
  \def\bibfnamefont#1{#1}\fi
\expandafter\ifx\csname citenamefont\endcsname\relax
  \def\citenamefont#1{#1}\fi
\expandafter\ifx\csname url\endcsname\relax
  \def\url#1{\texttt{#1}}\fi
\expandafter\ifx\csname urlprefix\endcsname\relax\def\urlprefix{URL }\fi
\providecommand{\bibinfo}[2]{#2}
\providecommand{\eprint}[2][]{\url{#2}}

\bibitem[{\citenamefont{Gasser}(2009)}]{gasser2009}
\bibinfo{author}{\bibfnamefont{U.}~\bibnamefont{Gasser}},
  \bibinfo{journal}{Journal of Physics: Condensed Matter}
  \textbf{\bibinfo{volume}{21}}, \bibinfo{pages}{203101}
  (\bibinfo{year}{2009}).

\bibitem[{\citenamefont{van Megen and Underwood}(1993)}]{vanmegen1993}
\bibinfo{author}{\bibfnamefont{W.}~\bibnamefont{van Megen}} \bibnamefont{and}
  \bibinfo{author}{\bibfnamefont{S.~M.} \bibnamefont{Underwood}},
  \bibinfo{journal}{Physical Review Letters} \textbf{\bibinfo{volume}{70}},
  \bibinfo{pages}{2766} (\bibinfo{year}{1993}).

\bibitem[{\citenamefont{van Megen and Underwood}(1994)}]{vanmegen1994}
\bibinfo{author}{\bibfnamefont{W.}~\bibnamefont{van Megen}} \bibnamefont{and}
  \bibinfo{author}{\bibfnamefont{S.~M.} \bibnamefont{Underwood}},
  \bibinfo{journal}{Physical Review E} \textbf{\bibinfo{volume}{49}},
  \bibinfo{pages}{4206} (\bibinfo{year}{1994}).

\bibitem[{\citenamefont{Sch\"ope et~al.}(2006)\citenamefont{Sch\"ope, Bryant,
  and van Megen}}]{schope2006}
\bibinfo{author}{\bibfnamefont{H.-J.} \bibnamefont{Sch\"ope}},
  \bibinfo{author}{\bibfnamefont{G.}~\bibnamefont{Bryant}}, \bibnamefont{and}
  \bibinfo{author}{\bibfnamefont{W.}~\bibnamefont{van Megen}},
  \bibinfo{journal}{Physical Review E} \textbf{\bibinfo{volume}{74}},
  \bibinfo{pages}{060401(R)} (\bibinfo{year}{2006}).

\bibitem[{\citenamefont{Sch\"ope et~al.}(2007)\citenamefont{Sch\"ope, Bryant,
  and van Megen}}]{schope2007}
\bibinfo{author}{\bibfnamefont{H.-J.} \bibnamefont{Sch\"ope}},
  \bibinfo{author}{\bibfnamefont{G.}~\bibnamefont{Bryant}}, \bibnamefont{and}
  \bibinfo{author}{\bibfnamefont{W.}~\bibnamefont{van Megen}},
  \bibinfo{journal}{Journal of Chemical Physics}
  \textbf{\bibinfo{volume}{127}}, \bibinfo{pages}{084505}
  (\bibinfo{year}{2007}).

\bibitem[{\citenamefont{Pham et~al.}(2002)\citenamefont{Pham, Puertas,
  Bergenholtz, Egelhaaf, Moussa\"id, Pusey, Schofield, Cates, Fuchs, and
  Poon}}]{pham2001}
\bibinfo{author}{\bibfnamefont{K.~N.} \bibnamefont{Pham}},
  \bibinfo{author}{\bibfnamefont{A.~M.} \bibnamefont{Puertas}},
  \bibinfo{author}{\bibfnamefont{J.}~\bibnamefont{Bergenholtz}},
  \bibinfo{author}{\bibfnamefont{S.~U.} \bibnamefont{Egelhaaf}},
  \bibinfo{author}{\bibfnamefont{A.}~\bibnamefont{Moussa\"id}},
  \bibinfo{author}{\bibfnamefont{P.~N.} \bibnamefont{Pusey}},
  \bibinfo{author}{\bibfnamefont{A.~B.} \bibnamefont{Schofield}},
  \bibinfo{author}{\bibfnamefont{M.~E.} \bibnamefont{Cates}},
  \bibinfo{author}{\bibfnamefont{M.}~\bibnamefont{Fuchs}}, \bibnamefont{and}
  \bibinfo{author}{\bibfnamefont{W.~C.~K.} \bibnamefont{Poon}},
  \bibinfo{journal}{Science} \textbf{\bibinfo{volume}{296}},
  \bibinfo{pages}{104} (\bibinfo{year}{2002}).

\bibitem[{\citenamefont{Pusey et~al.}(2009)\citenamefont{Pusey, Zaccarelli,
  Valeriani, Sanz, Poon, and Cates}}]{pusey2009}
\bibinfo{author}{\bibfnamefont{P.~N.} \bibnamefont{Pusey}},
  \bibinfo{author}{\bibfnamefont{E.}~\bibnamefont{Zaccarelli}},
  \bibinfo{author}{\bibfnamefont{C.}~\bibnamefont{Valeriani}},
  \bibinfo{author}{\bibfnamefont{E.}~\bibnamefont{Sanz}},
  \bibinfo{author}{\bibfnamefont{W.~C.~K.} \bibnamefont{Poon}},
  \bibnamefont{and} \bibinfo{author}{\bibfnamefont{M.~E.} \bibnamefont{Cates}},
  \bibinfo{journal}{Philosophical Transactions of the Royal Society A}
  \textbf{\bibinfo{volume}{367}}, \bibinfo{pages}{4993} (\bibinfo{year}{2009}).

\bibitem[{\citenamefont{Zaccarelli et~al.}(2015)\citenamefont{Zaccarelli,
  Liddle, and Poon}}]{zaccarelli2015}
\bibinfo{author}{\bibfnamefont{E.}~\bibnamefont{Zaccarelli}},
  \bibinfo{author}{\bibfnamefont{S.~M.} \bibnamefont{Liddle}},
  \bibnamefont{and} \bibinfo{author}{\bibfnamefont{W.~C.~K.}
  \bibnamefont{Poon}}, \bibinfo{journal}{Soft Matter}
  \textbf{\bibinfo{volume}{11}}, \bibinfo{pages}{324} (\bibinfo{year}{2015}).

\bibitem[{\citenamefont{Brambilla et~al.}(2009)\citenamefont{Brambilla,
  El~Masri, Pierno, Berthier, Cipelletti, Petekidis, and
  Schofield}}]{brambilla2009}
\bibinfo{author}{\bibfnamefont{G.}~\bibnamefont{Brambilla}},
  \bibinfo{author}{\bibfnamefont{D.}~\bibnamefont{El~Masri}},
  \bibinfo{author}{\bibfnamefont{M.}~\bibnamefont{Pierno}},
  \bibinfo{author}{\bibfnamefont{L.}~\bibnamefont{Berthier}},
  \bibinfo{author}{\bibfnamefont{L.}~\bibnamefont{Cipelletti}},
  \bibinfo{author}{\bibfnamefont{G.}~\bibnamefont{Petekidis}},
  \bibnamefont{and} \bibinfo{author}{\bibfnamefont{A.~B.}
  \bibnamefont{Schofield}}, \bibinfo{journal}{Physical Review Letters}
  \textbf{\bibinfo{volume}{102}}, \bibinfo{pages}{085703}
  (\bibinfo{year}{2009}).

\bibitem[{\citenamefont{Klochko et~al.}(2020)\citenamefont{Klochko, Baschnagel,
  Wittmer, Benzerara, Ruscher, and Semenov}}]{klochko2020}
\bibinfo{author}{\bibfnamefont{L.}~\bibnamefont{Klochko}},
  \bibinfo{author}{\bibfnamefont{J.}~\bibnamefont{Baschnagel}},
  \bibinfo{author}{\bibfnamefont{J.~P.} \bibnamefont{Wittmer}},
  \bibinfo{author}{\bibfnamefont{O.}~\bibnamefont{Benzerara}},
  \bibinfo{author}{\bibfnamefont{C.}~\bibnamefont{Ruscher}}, \bibnamefont{and}
  \bibinfo{author}{\bibfnamefont{A.~N.} \bibnamefont{Semenov}},
  \bibinfo{journal}{Physical Review E} \textbf{\bibinfo{volume}{102}},
  \bibinfo{pages}{042611} (\bibinfo{year}{2020}).

\bibitem[{\citenamefont{Leocmach et~al.}(2013)\citenamefont{Leocmach, Russo,
  and Tanaka}}]{leocmach2013}
\bibinfo{author}{\bibfnamefont{M.}~\bibnamefont{Leocmach}},
  \bibinfo{author}{\bibfnamefont{J.}~\bibnamefont{Russo}}, \bibnamefont{and}
  \bibinfo{author}{\bibfnamefont{H.}~\bibnamefont{Tanaka}},
  \bibinfo{journal}{Journal of Chemical Physics}
  \textbf{\bibinfo{volume}{138}}, \bibinfo{pages}{12A536}
  (\bibinfo{year}{2013}).

\bibitem[{\citenamefont{Ingebrigtsen and Tanaka}(2015)}]{ingebrigtsen2015}
\bibinfo{author}{\bibfnamefont{T.~S.} \bibnamefont{Ingebrigtsen}}
  \bibnamefont{and} \bibinfo{author}{\bibfnamefont{H.}~\bibnamefont{Tanaka}},
  \bibinfo{journal}{Journal of Physical Chemistry B}
  \textbf{\bibinfo{volume}{119}}, \bibinfo{pages}{11052}
  (\bibinfo{year}{2015}).

\bibitem[{\citenamefont{Ingebrigtsen et~al.}(2021)\citenamefont{Ingebrigtsen,
  Schr\o{}der, and Dyre}}]{ingebrigtsen2021}
\bibinfo{author}{\bibfnamefont{T.~S.} \bibnamefont{Ingebrigtsen}},
  \bibinfo{author}{\bibfnamefont{T.~B.} \bibnamefont{Schr\o{}der}},
  \bibnamefont{and} \bibinfo{author}{\bibfnamefont{J.~C.} \bibnamefont{Dyre}},
  \bibinfo{journal}{Journal of Physical Chemistry B}
  \textbf{\bibinfo{volume}{125}}, \bibinfo{pages}{317} (\bibinfo{year}{2021}).

\bibitem[{\citenamefont{Ninarello et~al.}(2017)\citenamefont{Ninarello,
  Berthier, and Coslovich}}]{ninarello2017}
\bibinfo{author}{\bibfnamefont{A.}~\bibnamefont{Ninarello}},
  \bibinfo{author}{\bibfnamefont{L.}~\bibnamefont{Berthier}}, \bibnamefont{and}
  \bibinfo{author}{\bibfnamefont{D.}~\bibnamefont{Coslovich}},
  \bibinfo{journal}{Physical Review X} \textbf{\bibinfo{volume}{7}},
  \bibinfo{pages}{021039} (\bibinfo{year}{2017}).

\bibitem[{\citenamefont{Guiselin
  et~al.}(2020{\natexlab{a}})\citenamefont{Guiselin, Tarjus, and
  Berthier}}]{guiselin2020overlap}
\bibinfo{author}{\bibfnamefont{B.}~\bibnamefont{Guiselin}},
  \bibinfo{author}{\bibfnamefont{G.}~\bibnamefont{Tarjus}}, \bibnamefont{and}
  \bibinfo{author}{\bibfnamefont{L.}~\bibnamefont{Berthier}},
  \bibinfo{journal}{The Journal of Chemical Physics}
  \textbf{\bibinfo{volume}{153}}, \bibinfo{pages}{224502}
  (\bibinfo{year}{2020}{\natexlab{a}}).

\bibitem[{\citenamefont{Vaibhav et~al.}(2022)\citenamefont{Vaibhav, Horbach,
  and Chaudhuri}}]{vaibhav2022}
\bibinfo{author}{\bibfnamefont{V.}~\bibnamefont{Vaibhav}},
  \bibinfo{author}{\bibfnamefont{J.}~\bibnamefont{Horbach}}, \bibnamefont{and}
  \bibinfo{author}{\bibfnamefont{P.}~\bibnamefont{Chaudhuri}},
  \bibinfo{journal}{Journal of Chemical Physics}
  \textbf{\bibinfo{volume}{156}}, \bibinfo{pages}{244501}
  (\bibinfo{year}{2022}).

\bibitem[{\citenamefont{Lamp et~al.}(2022)\citenamefont{Lamp, K\"uchler, and
  Horbach}}]{lamp2022}
\bibinfo{author}{\bibfnamefont{K.}~\bibnamefont{Lamp}},
  \bibinfo{author}{\bibfnamefont{N.}~\bibnamefont{K\"uchler}},
  \bibnamefont{and} \bibinfo{author}{\bibfnamefont{J.}~\bibnamefont{Horbach}},
  \bibinfo{journal}{Journal of Chemical Physics}
  \textbf{\bibinfo{volume}{157}}, \bibinfo{pages}{034501}
  (\bibinfo{year}{2022}).

\bibitem[{\citenamefont{Voigtmann and Horbach}(2009)}]{voigtmann2009}
\bibinfo{author}{\bibfnamefont{T.}~\bibnamefont{Voigtmann}} \bibnamefont{and}
  \bibinfo{author}{\bibfnamefont{J.}~\bibnamefont{Horbach}},
  \bibinfo{journal}{Physical Review Letters} \textbf{\bibinfo{volume}{103}},
  \bibinfo{pages}{205901} (\bibinfo{year}{2009}).

\bibitem[{\citenamefont{Weysser et~al.}(2010)\citenamefont{Weysser, Puertas,
  Fuchs, and Voigtmann}}]{weysser2010}
\bibinfo{author}{\bibfnamefont{F.}~\bibnamefont{Weysser}},
  \bibinfo{author}{\bibfnamefont{A.~M.} \bibnamefont{Puertas}},
  \bibinfo{author}{\bibfnamefont{M.}~\bibnamefont{Fuchs}}, \bibnamefont{and}
  \bibinfo{author}{\bibfnamefont{T.}~\bibnamefont{Voigtmann}},
  \bibinfo{journal}{Physical Review E} \textbf{\bibinfo{volume}{82}},
  \bibinfo{pages}{011504} (\bibinfo{year}{2010}).

\bibitem[{\citenamefont{Santen and Krauth}(2001)}]{santen2001liquid}
\bibinfo{author}{\bibfnamefont{L.}~\bibnamefont{Santen}} \bibnamefont{and}
  \bibinfo{author}{\bibfnamefont{W.}~\bibnamefont{Krauth}},
  \bibinfo{journal}{arXiv preprint cond-mat/0107459}  (\bibinfo{year}{2001}).

\bibitem[{\citenamefont{Tsai et~al.}(1978)\citenamefont{Tsai, Abraham, and
  Pound}}]{tsai1978structure}
\bibinfo{author}{\bibfnamefont{N.-H.} \bibnamefont{Tsai}},
  \bibinfo{author}{\bibfnamefont{F.~F.} \bibnamefont{Abraham}},
  \bibnamefont{and} \bibinfo{author}{\bibfnamefont{G.}~\bibnamefont{Pound}},
  \bibinfo{journal}{Surface Science} \textbf{\bibinfo{volume}{77}},
  \bibinfo{pages}{465} (\bibinfo{year}{1978}).

\bibitem[{\citenamefont{Grigera and Parisi}(2001)}]{grigera2001fast}
\bibinfo{author}{\bibfnamefont{T.~S.} \bibnamefont{Grigera}} \bibnamefont{and}
  \bibinfo{author}{\bibfnamefont{G.}~\bibnamefont{Parisi}},
  \bibinfo{journal}{Physical Review E} \textbf{\bibinfo{volume}{63}},
  \bibinfo{pages}{045102(R)} (\bibinfo{year}{2001}).

\bibitem[{\citenamefont{Guiselin
  et~al.}(2020{\natexlab{b}})\citenamefont{Guiselin, Berthier, and
  Tarjus}}]{RFIM_in_glassforming_liquid}
\bibinfo{author}{\bibfnamefont{B.}~\bibnamefont{Guiselin}},
  \bibinfo{author}{\bibfnamefont{L.}~\bibnamefont{Berthier}}, \bibnamefont{and}
  \bibinfo{author}{\bibfnamefont{G.}~\bibnamefont{Tarjus}},
  \bibinfo{journal}{Physical Review E} \textbf{\bibinfo{volume}{102}},
  \bibinfo{pages}{042129} (\bibinfo{year}{2020}{\natexlab{b}}).

\bibitem[{\citenamefont{Berthier et~al.}(2019)\citenamefont{Berthier, Flenner,
  Fullerton, Scalliet, and Singh}}]{berthier2019efficient}
\bibinfo{author}{\bibfnamefont{L.}~\bibnamefont{Berthier}},
  \bibinfo{author}{\bibfnamefont{E.}~\bibnamefont{Flenner}},
  \bibinfo{author}{\bibfnamefont{C.~J.} \bibnamefont{Fullerton}},
  \bibinfo{author}{\bibfnamefont{C.}~\bibnamefont{Scalliet}}, \bibnamefont{and}
  \bibinfo{author}{\bibfnamefont{M.}~\bibnamefont{Singh}},
  \bibinfo{journal}{Journal of Statistical Mechanics: Theory and Experiment}
  \textbf{\bibinfo{volume}{2019}}, \bibinfo{pages}{064004}
  (\bibinfo{year}{2019}).

\bibitem[{\citenamefont{Koopman and
  Lowe}(2006)}]{Lowe_Andersen_T_different_masses}
\bibinfo{author}{\bibfnamefont{E.}~\bibnamefont{Koopman}} \bibnamefont{and}
  \bibinfo{author}{\bibfnamefont{C.}~\bibnamefont{Lowe}}, \bibinfo{journal}{The
  Journal of chemical physics} \textbf{\bibinfo{volume}{124}},
  \bibinfo{pages}{204103} (\bibinfo{year}{2006}).

\bibitem[{\citenamefont{Matsumoto and Nishimura}(1998)}]{matsumoto1998mersenne}
\bibinfo{author}{\bibfnamefont{M.}~\bibnamefont{Matsumoto}} \bibnamefont{and}
  \bibinfo{author}{\bibfnamefont{T.}~\bibnamefont{Nishimura}},
  \bibinfo{journal}{ACM Transactions on Modeling and Computer Simulation
  (TOMACS)} \textbf{\bibinfo{volume}{8}}, \bibinfo{pages}{3}
  (\bibinfo{year}{1998}).

\bibitem[{\citenamefont{Efron}(1992)}]{efron1992bootstrap}
\bibinfo{author}{\bibfnamefont{B.}~\bibnamefont{Efron}}, in
  \emph{\bibinfo{booktitle}{Breakthroughs in statistics}}
  (\bibinfo{publisher}{Springer}, \bibinfo{year}{1992}), pp.
  \bibinfo{pages}{569--593}.

\bibitem[{\citenamefont{Lebowitz et~al.}(1967)\citenamefont{Lebowitz, Percus,
  and Verlet}}]{lebowitz1967ensemble}
\bibinfo{author}{\bibfnamefont{J.}~\bibnamefont{Lebowitz}},
  \bibinfo{author}{\bibfnamefont{J.}~\bibnamefont{Percus}}, \bibnamefont{and}
  \bibinfo{author}{\bibfnamefont{L.}~\bibnamefont{Verlet}},
  \bibinfo{journal}{Physical Review} \textbf{\bibinfo{volume}{153}},
  \bibinfo{pages}{250} (\bibinfo{year}{1967}).

\bibitem[{\citenamefont{Scheidler et~al.}(2001)\citenamefont{Scheidler, Kob,
  Latz, Horbach, and Binder}}]{scheidler2001}
\bibinfo{author}{\bibfnamefont{P.}~\bibnamefont{Scheidler}},
  \bibinfo{author}{\bibfnamefont{W.}~\bibnamefont{Kob}},
  \bibinfo{author}{\bibfnamefont{A.}~\bibnamefont{Latz}},
  \bibinfo{author}{\bibfnamefont{J.}~\bibnamefont{Horbach}}, \bibnamefont{and}
  \bibinfo{author}{\bibfnamefont{K.}~\bibnamefont{Binder}},
  \bibinfo{journal}{Physical Review B} \textbf{\bibinfo{volume}{63}},
  \bibinfo{pages}{104204} (\bibinfo{year}{2001}).

\bibitem[{\citenamefont{Chandler et~al.}(2006)\citenamefont{Chandler, Garrahan,
  Jack, Maibaum, and Pan}}]{chandler2006}
\bibinfo{author}{\bibfnamefont{D.}~\bibnamefont{Chandler}},
  \bibinfo{author}{\bibfnamefont{J.~P.} \bibnamefont{Garrahan}},
  \bibinfo{author}{\bibfnamefont{R.~L.} \bibnamefont{Jack}},
  \bibinfo{author}{\bibfnamefont{L.}~\bibnamefont{Maibaum}}, \bibnamefont{and}
  \bibinfo{author}{\bibfnamefont{A.~C.} \bibnamefont{Pan}},
  \bibinfo{journal}{Phys. Rev. E} \textbf{\bibinfo{volume}{74}},
  \bibinfo{pages}{051501} (\bibinfo{year}{2006}),
  \urlprefix\url{https://link.aps.org/doi/10.1103/PhysRevE.74.051501}.

\bibitem[{\citenamefont{Cavagna}(2009)}]{cavagna2009}
\bibinfo{author}{\bibfnamefont{A.}~\bibnamefont{Cavagna}},
  \bibinfo{journal}{Physics Reports} \textbf{\bibinfo{volume}{476}},
  \bibinfo{pages}{51} (\bibinfo{year}{2009}).

\bibitem[{\citenamefont{Chung}(1974)}]{chung1974}
\bibinfo{author}{\bibfnamefont{K.~L.} \bibnamefont{Chung}},
  \emph{\bibinfo{title}{A Course in Probability Theory}}
  (\bibinfo{publisher}{Academic Press}, \bibinfo{address}{New York},
  \bibinfo{year}{1974}).

\bibitem[{\citenamefont{Popoviciu}(1935)}]{popoviciu1935equations}
\bibinfo{author}{\bibfnamefont{T.}~\bibnamefont{Popoviciu}},
  \bibinfo{journal}{Mathematica} \textbf{\bibinfo{volume}{9}},
  \bibinfo{pages}{20} (\bibinfo{year}{1935}).

\bibitem[{\citenamefont{Lerner}(2019)}]{lerner2019}
\bibinfo{author}{\bibfnamefont{E.}~\bibnamefont{Lerner}},
  \bibinfo{journal}{Journal of Non-Crystalline Solids}
  \textbf{\bibinfo{volume}{522}}, \bibinfo{pages}{119570}
  (\bibinfo{year}{2019}), ISSN \bibinfo{issn}{0022-3093},
  \urlprefix\url{https://www.sciencedirect.com/science/article/pii/S0022309319304417}.

\bibitem[{\citenamefont{Rainone et~al.}(2020)\citenamefont{Rainone,
  Bouchbinder, and Lerner}}]{rainone2020}
\bibinfo{author}{\bibfnamefont{C.}~\bibnamefont{Rainone}},
  \bibinfo{author}{\bibfnamefont{E.}~\bibnamefont{Bouchbinder}},
  \bibnamefont{and} \bibinfo{author}{\bibfnamefont{E.}~\bibnamefont{Lerner}},
  \bibinfo{journal}{Proceedings of the National Academy of Sciences}
  \textbf{\bibinfo{volume}{117}}, \bibinfo{pages}{5228} (\bibinfo{year}{2020}),
  \eprint{https://www.pnas.org/doi/pdf/10.1073/pnas.1919958117},
  \urlprefix\url{https://www.pnas.org/doi/abs/10.1073/pnas.1919958117}.

\end{thebibliography}
\end{document}